\documentclass[%
 reprint,
 superscriptaddress,
 amsmath,amssymb,
 aip,
 jcp,
floatfix,
]{revtex4-2}

\usepackage{color,graphics}
\usepackage{graphicx}
\usepackage[sort&compress]{natbib}
\usepackage{hhline}
\usepackage{soul}
\usepackage{url}
\usepackage{hyperref}
\usepackage{graphicx}% Include figure files
\usepackage{dcolumn}% Align table columns on decimal point
\usepackage{bm}% bold math
\usepackage[utf8]{inputenc}
\usepackage[T1]{fontenc}
\usepackage{mathptmx}
\usepackage{etoolbox}

\usepackage[usenames,dvipsnames]{xcolor}

\makeatletter
\def\@email#1#2{%
 \endgroup
 \patchcmd{\titleblock@produce}
  {\frontmatter@RRAPformat}
  {\frontmatter@RRAPformat{\produce@RRAP{*#1\href{mailto:#2}{#2}}}\frontmatter@RRAPformat}
  {}{}
}%
\makeatother
\begin{document}

\preprint{%AIP/123-QED
}

\title[]{
Grand-canonical molecular dynamics simulations powered by a hybrid 4D nonequilibrium MD/MC method: Implementation in LAMMPS and applications to electrolyte solutions}
% Force line breaks with \\
\author{Jeongmin Kim}
\address
{Sorbonne Universit\'{e}, CNRS, Physico-chimie des \'{E}lectrolytes et Nanosystem\`{e}s Interfaciaux, PHENIX, F-75005 Paris, France}
\author{Luc Belloni}
\affiliation
{LIONS, NIMBE, CEA, CNRS, Universit\'{e} Paris-Saclay, 91191-Gif-sur-Yvette, France}
\author{Benjamin Rotenberg}
\email{benjamin.rotenberg@sorbonne-universite.fr}
\address
{Sorbonne Universit\'{e}, CNRS, Physico-chimie des \'{E}lectrolytes et Nanosystem\`{e}s Interfaciaux, PHENIX, F-75005 Paris, France}
\affiliation{R\'eseau sur le Stockage Electrochimique de l'Energie (RS2E), FR CNRS 3459, 80039 Amiens Cedex, France}

\date{\today}% It is always \today, today,

\begin{abstract}
Molecular simulations in an open environment, involving ion exchange, are necessary to study various systems, from biosystems to confined electrolytes. However, grand-canonical simulations are often computationally demanding in condensed phases. A promising method (L. Belloni, J. Chem. Phys., 2019), one of the hybrid nonequilibrium molecular dynamics/Monte Carlo algorithms, was recently developed, which enables efficient computation of fluctuating number or charge density in dense fluids or ionic solutions. This method facilitates the exchange through an auxiliary dimension, orthogonal to all physical dimensions, by reducing initial steric and electrostatic clashes in three-dimensional systems. Here, we report the implementation of the method in LAMMPS with a Python interface, allowing facile access to grand-canonical molecular dynamics (GCMD) simulations with massively parallelized computation.  We validate our implementation with two electrolytes, including a model Lennard-Jones electrolyte similar to a restricted primitive model and aqueous solutions. We find that electrostatic interactions play a crucial role in the overall efficiency due to their long-range nature, particularly for water or ion-pair exchange in aqueous solutions. With properly screened electrostatic interactions and bias-based methods, our approach enhances the efficiency of salt-pair exchange in Lennard-Jones electrolytes by approximately four orders of magnitude, compared to conventional grand-canonical Monte Carlo. Furthermore, the acceptance rate of NaCl-pair exchange in aqueous solutions at moderate concentrations reaches about 3 $\%$ at the maximum efficiency. 
\end{abstract}

\maketitle

\section{Introduction}
Molecular simulations in an open environment are necessary to study various systems, including electrolytes \cite{valeau1980primitive,belloni2019non}, porous materials \cite{smit1995grand}, and biosystems \cite{stern2007molecular,michael2020hybrid}. Fluctuations in an open system allow for calculating thermodynamic derivatives, \textit{i.e.}, responses of the system upon a perturbation.
For example, the osmotic compressibility $\chi_{osmotic}$ is related to the fluctuation in salt density $\rho_{\text{salt}}$, as follows:
\begin{equation}
\begin{split}\label{osmotic}
\chi_{osmotic}
& =k_BT\bigg(\frac{\partial\rho_{\text{salt}}}{\partial P_{osmotic}}\bigg)_{T} \\
& =\langle V\rangle
\frac{\langle( \rho_{\text{salt}}-\langle\rho_{\text{salt}}\rangle)^2\rangle}{\langle\rho_{\text{salt}}\rangle}
\end{split}
\end{equation}
with $\langle\cdots\rangle$ is the ensemble average in an open system in temperature $T$ whose average volume is $\langle V\rangle$, and $k_B$ Boltzmann constant.
For sufficiently dilute 1:1 electrolytes, this leads to the ideal (non-interacting) result $\chi_{osmotic}=0.5$. Thermodynamic derivatives can also be computed via so-called Kirkwood-Buff integrals \cite{kirkwood1951statistical,kusalik1987thermodynamic,cheng2022computing} in both fluctuating and constant particle-number simulations.
One of the advantages in grand-canonical ensembles is that no further treatment (\textit{e.g.}, the so-called finite-size corrections) is needed to the Kirkwood-Buff integrals \cite{belloni2019non}, which is essential in constant particle-number simulations.

And yet, grand-canonical (GC) simulations are often computationally demanding, particularly with explicit solvents in condensed phases where almost all trial insertion or deletion moves would be rejected.
Several advanced Monte Carlo methods for the efficient exchange have been developed, including a cavity-bias method \cite{mezei1980cavity}, a configuration-biased method \cite{shelley1994configuration,shelley1995configuration,smit1995grand}, a Boltzmann-bias method \cite{garberoglio2008boltzmann}, a continuous fraction component method \cite{shi2007continuous}, identity exchange methods \cite{panagiotopoulos1989exact,soroush2018molecular,fathizadeh2018mixed}, and hybrid methods \cite{duane1987hybrid,mehlig1992hybrid,boinepalli2003grand,stern2007molecular,chen2014efficient,radak2017constant,ross2018biomolecular,prokhorenko2018large,nilmeier2011nonequilibrium,belloni2019non}.
However, the computational efficiency is still limited, leaving an obstacle to simulating systems like ionic solutions due to long-range electrostatic interactions and strong short-range steric repulsion.
Thermodynamic extrapolation~\cite{schneck2011simple,schneck2012hydration,schlaich_water_2016} is an alternative approach which does not require particle exchange, but allows to determine  iteratively the appropriate number of particles (or volume) corresponding to a given chemical potential, based on the difference between the measured and target value. Yet, its computational cost is also significant and it does not correspond exactly to the grand-canonical ensemble (since the number of particles does not fluctuate).

Hybrid nonequilibrium MD/MC methods \cite{duane1987hybrid,mehlig1992hybrid,boinepalli2003grand,stern2007molecular,chen2014efficient,ross2018biomolecular,nilmeier2011nonequilibrium,belloni2019non} are an attractive means to enhance efficiency in grand-canonical simulations with a proper design of a path for the exchange.
The hybrid method generates a proposed transition from one configuration to another via MD propagation during a finite time instead of an instantaneous (infinitely fast) switching.
A trial move, thus, is not instantaneous or local in space, \textit{i.e.}, all the particles could relax in principle.
During the slow switching, a time-dependent nonequilibrium Hamiltonian governs the system's time evolution.
After such a finite switching, a Metropolis criterion is applied, which involves the nonequilibrium (NE) work, including the changes of potential and kinetic energies, to decide whether or not a trial move is accepted.

One of the successful methods is a MC osmostat \cite{ross2018biomolecular,izarra2023alchemical} to simulate a system in a fluctuating ionic environment that could differ from the bulk in a semigrand canonical ensemble.
In their approach, two randomly chosen water molecules are "transmuted" to a pair of Na$^+$ and Cl$^-$ or vice versa, ensuring the total charge neutrality.
Such exchange between water and ions is done slowly via nonequilibrium candidate MC (NCMC) \cite{nilmeier2011nonequilibrium}, providing a finite time for the collective relaxation.
The optimal efficiency is determined by the acceptance rate per the time spent in a NCMC switching, although the acceptance rate generally increases with the increasing amount of time in the MD propagation.
With a proper NE protocol, the NCMC achieved the acceptance rate of $~\sim20\%$ with a 20 ps long NCMC protocol at maximum efficiency.

Designing a nonequilibrium Hamiltonian in a hybrid NEMD/MC method is arbitrary, and a well-designed path should enhance the acceptance rate.
Such flexibility helps the hybrid method \cite{nilmeier2011nonequilibrium,kurut2017driving,suh2018enhanced,fathizadeh2018mixed,sasmal2020sampling} sample a rare barrier crossing event in a rugged free energy surface, showing a faster convergence than conventional equilibrium MD.
The lowered free energy barrier with a Hamiltonian perturbation can facilitate the dynamics of a particular part of interest (usually slow modes).
A perturbation scheme during NEMD to construct a NE Hamiltonian can be chosen to deform (flatten) a potential energy surface \cite{suh2018enhanced}, including, the accelerated MD \cite{voter1997hyperdynamics}, the replica-exchange with solute tempering \cite{wang2011replica}, and alchemical mixing \cite{fathizadeh2018mixed}.
Thus, choosing such a coupling method is crucial in determining the sampling efficiency.

In this work, we report the implementation of a particular hybrid NEMD/MC method, called "H4D" \cite{belloni2019non}, in LAMMPS \cite{plimpton1995fast} and its applications to ionic solutions in a wide range of salinity.
As was well illustrated in the original paper \cite{belloni2019non}, the idea of H4D is to utilize an auxiliary, non-physical dimension to facilitate the exchange, 
which is orthogonal to all other physical dimensions. 
For example, the "vertical" axis is the third dimension in two-dimensional systems or the fourth dimension in three-dimensional systems.
A NE Hamiltonian for the exchange considers a time-dependent "altitude" schedule along the vertical dimension.
The potential energy ($U$) becomes a function of a time-dependent altitude, $w(t)$: $U=U(\{\vec{r}\},w(t))$ with position vectors ($\vec{r}$) of all atoms.
For a trial insertion move, a particle to exchange starts at a non-zero altitude, which will finally land at a zero altitude.
Further, the altitude schedule is a critical factor for efficiency, which determines how fast a trial insertion or deletion is, \textit{i.e.}, the vertical velocity ($v_f$).
That is, in H4D, a particle to exchange "flies" through the vertical dimension at a particular vertical velocity.
Such a flying scheme alleviates the significant steric repulsion in the initial steps of NEMD, increasing the exchange's acceptance rate.

The H4D method is conceptually simple \cite{belloni2019non} and works well to reduce steric clashes with the 4D distance considering the altitude for a 3D system during NEMD.
However, ion-pair exchange via the H4D is still challenging because it introduces a strong perturbation with charge-charge and charge-dipole interactions.
It was previously shown that NaCl-pair exchange in aqueous solutions resulted in a significantly lower acceptance rate than water exchange, even with an order of magnitude longer NEMD \cite{belloni2019non}.
In this work, we show that designing a NE Hamiltonian with screened electrostatic interactions can significantly enhance the acceptance rate.
In particular, an altitude-based screening function benefits the efficiency of NaCl-pair exchange in aqueous solutions by reducing the Coulomb coupling between flying and non-flying particles in real space.
We also find that the Coulomb interactions, reduced in their magnitude, still require a long cut-off distance that partitions the short- and long-range contributions.
Our findings assure the significance of the long-range nature of electrostatic interactions in ionic solutions.

The paper is organized as follows.
Section~\ref{methods} discusses our method and implementation of grand-canonical molecular dynamics (GCMD) simulations, including the basics of H4D, the techniques to enhance efficiency, and a brief illustration of our implementation in LAMMPS that allows for the massively parallelized computations.
Some of those were briefly discussed in the original paper \cite{belloni2019non}.
Section~\ref{methods} also presents model electrolytes for applications of our implementation and a chemical potential calculation using H4D.
Section~\ref{results} discusses the results of GCMD simulations using H4D, including the efficiency of solvent or ion-pair exchange, as well as number fluctuation and its convergence, followed by Conclusions Section.

\section{Simulation Methods and model systems}\label{methods}
As in Figure~\ref{h4d}, GCMD simulations are composed of equilibrium MD to sample configurations, NEMD to propose a trial move for particle exchange, and MC to decide whether or not the trial move is accepted, following the Metropolis acceptance rule. 
We have implemented a hybrid NEMD/MC (H4D) method \cite{belloni2019non} in LAMMPS with a Python interface, which is efficient for particle exchange in dense fluids or electrolytes. 
The following section discusses details of a H4D method and systems to validate our implementation.

\subsection{A hybrid 4D NEMD/MC scheme for exchange}
\begin {figure}[htbp]
\includegraphics [width=3.5in] {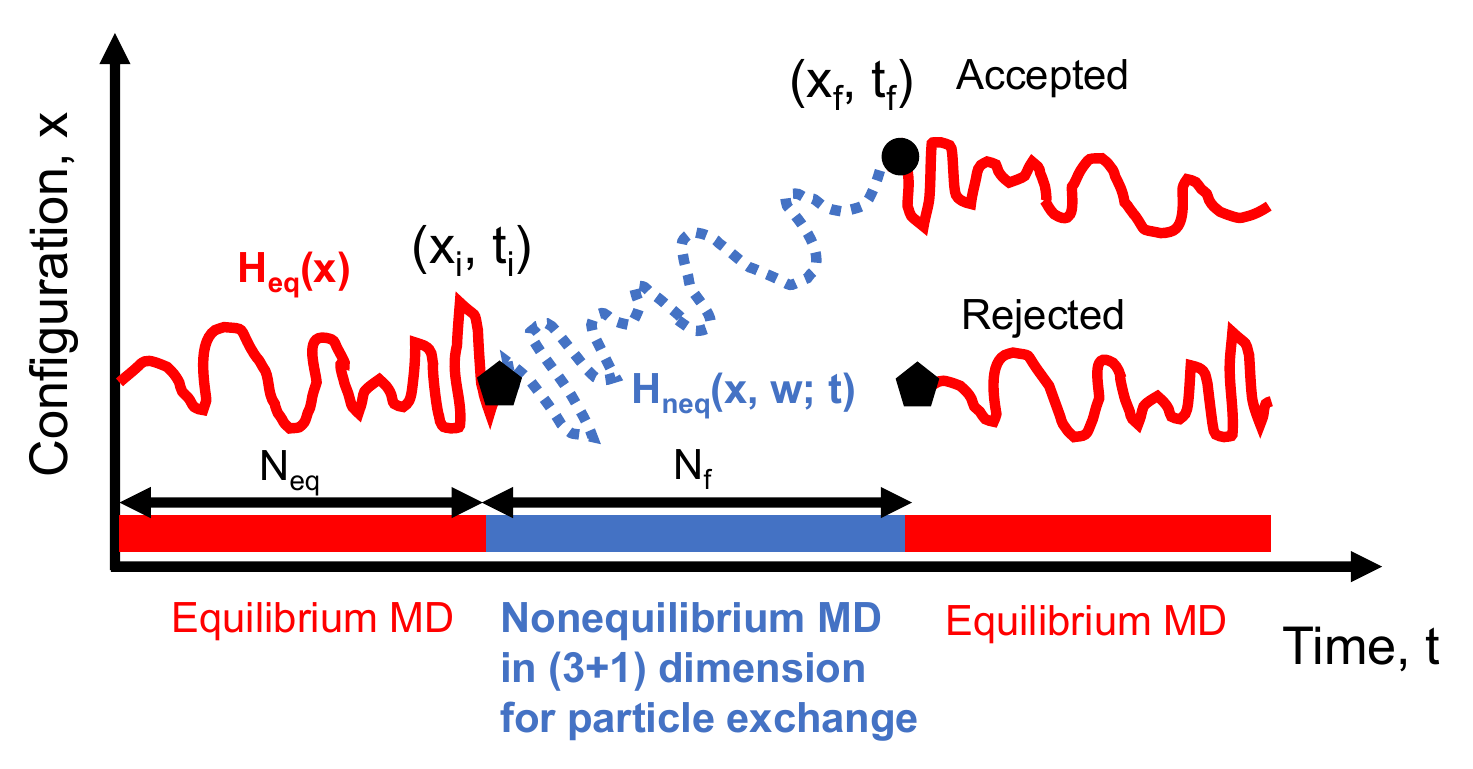}
\caption{Schematic illustration of a grand-canonical molecular dynamics (GCMD) simulation with H4D. 
Equilibrium MD is in red, and nonequilibrium MD in 4D, including a vertical dimension, for particle exchange is in blue.
$N_{eq}$ and $N_f$ are the number of steps in equilibrium and nonequilibrium MD, respectively.
$H_{eq}$ is an equilibrium Hamiltonian, and $H_{neq}$ is a nonequilibrium Hamiltonian with time-dependent "altitude", $w$.
Regardless of the ensemble chosen in equilibrium MD, the time evolution during nonequilibrium MD is done in microcanonical ensemble.
}
\label{h4d}
\end{figure}

In this section, we discuss a hybrid 4D NEMD/MC method, including the time evolution of NEMD, and a Metropolis scheme that determines whether a trial exchange move is accepted or rejected.

\textbf{Constant-velocity altitude protocol in 4D NEMD.}
The essence of H4D is to utilize the fourth dimension, called "vertical" dimension, to facilitate particle exchange, which is orthogonal to all other physical dimensions.
That is, a set of particles are inserted or removed through the "vertical" axis.
The key parameters of H4D include the maximum altitude ($w_{max}$) and vertical velocity ($v_f(t)$), both of which are crucial to determine the efficiency of particle exchange.
Both parameters also determine the time-dependent altitude, $w(t)$, of a so-called flying particle to exchange, which is changed in a controlled manner during NEMD.
We note that all other particles except flying ones are forced to remain in 3D, \textit{i.e.}, $w(t)=0$ at all $t$.
There are several ways to define $w(t)$.
In this work, we use a constant-velocity $w(t)$, following Ref.~\onlinecite{belloni2019non} (See Section~\ref{sec:SI_altitude} of SI for other possible altitude schedules).
\begin{equation} \label{const-vel}
w(t)=v_f\cdot(t-t_i) + w_{max},
\end{equation}
where $t\in[t_i,t_f]$, and $v_f$ is a constant velocity of the altitude whose sign depends on whether a trial move is for addition or for removal.
The altitude should satisfy the following boundary conditions.
For a trial insertion move, $w(t_i)=w_{max}$ and $w(t_f)=0$, while for a trial removal move, $w(t_i)=0$ and $w(t_f)=w_{max}$.
By doing so, the insertion and removal moves have a symmetric pre-determined altitude schedule to satisfy the detailed balance condition.
In the case of the exchange of multiple particles, their altitude does not need to be the same as long as their altitude schedule is controlled and reversible (See Section~\ref{sec:SI_altitude} of SI).
However, one needs to take special care when using altitude-dependent bias techniques discussed in Section~\ref{subsection:trick}.

\textbf{Time evolution of NEMD.}
As in other hybrid methods \cite{mehlig1992hybrid,guo2018hybrid,belloni2019non}, we chose a deterministic integrator in the \textit{NVE} ensemble with no heat exchange with the reservoir, which is time-reversible, and volume-preserving \cite{mehlig1992hybrid}.
Then, a trial transition proposed by H4D from one state to other via a Hamiltonian is solely determined by the preparation of all momenta. %ith a deterministic propagator during NEMD in the microcanonical ensemble.
The prepared random momenta of flying particles should follow the Maxwell-Boltzmann distribution, satisfying the equipartition theorem; \textit{i.e.}, $\langle v^2_x\rangle=k_BT/m$ for a linear momentum along $x$ axis of a single particle of mass $m$.
For the systems with rigid molecules such as SPC/E water, particular care must be taken in two ways: the proper sampling of random initial velocities and the time-reversible evolution.
First, we draw random initial velocities of flying rigid water that faithfully obeys the equipartition theorem.
This is achieved by initializing the orientation and momentum of flying water molecules, following the procedure in Ref~\onlinecite{palmer2018comment}: the random velocities of the flying water are drawn in the molecular frame, and transformed to the laboratory frame.
Second, we use a quaternion-based propagator \cite{miller2002symplectic}, which is symplectic and reversible, for all rigid water molecules, including a flying one.
We note that the NEMD could be performed in the \textit{NVT} ensemble as well, with exchange of heat with a reservoir \cite{chen2014efficient,belloni2019non}. 
However, Ref.~\onlinecite{belloni2019non} found no improvement in the acceptance probability.

Since we evolve the systems in the equilibrium phase using MD as well (Fig.~\ref{h4d}), the momentum reversal scheme for the detailed balance should be considered \cite{nilmeier2011nonequilibrium,chen2014efficient}.
We chose a method of symmetric two-end momentum reversal \cite{chen2014efficient} for a fast decorrelation, instead of a one-end reversal scheme \cite{nilmeier2011nonequilibrium}, so the momenta of all particles will be reversed by a chance of one-half on average at both ends of NEMD for every trial move.

\textbf{Metropolis scheme of the acceptance rule.}
As in other hybrid methods \cite{mehlig1992hybrid,guo2018hybrid,belloni2019non}, the Metropolis acceptance probability, $f_{i\rightarrow j}$, of H4D includes an additional factor that is related to the time evolution of a system, \textit{i.e.}, the transition from a state $i$ to the other state $j$ during MD.
For a deterministic propagator during NEMD, this factor is solely determined by the way to prepare all momenta of the system, including "flying" particles.
In a conventional hybrid MC \cite{mehlig1992hybrid} with constant number of particles, the probability density to generate state $j$ from state $i$ is $\alpha(i\rightarrow j)\propto \exp(-\beta\sum_n {\vec{p}_n}^2/2m_n)=\exp(-\beta K_i)$, where $n$ is a particle index, and $\vec{p}_n$, and $m_n$ are the momentum vector and mass of $n^{th}$ particle, respectively. $K_i$ is the kinetic energy of state $i$.

Here, we present the Metropolis acceptance probability for exchange of an ion pair, which can be easily translated to the one for exchange of single flying particle (e.g., water molecules).
For a trial ion-pair deletion move ($i,N_{salt}+1 \rightarrow j,N_{salt}$),
\begin{equation}\label{trial_del}
    % \alpha_{i,N_{salt}+1 \rightarrow j,N_{salt}}
   \begin{split}
   &\alpha(i,N_{salt}+1 \rightarrow j,N_{salt})\\
     &= \bigg(\frac{1}{N_{salt}+1}\bigg)^2 
    \frac{\exp(-\beta K_i)}{\mathcal{N}_k},       
   \end{split}
    %\frac{1}{\mathcal{N}_k(\beta, N_{salt}+1)},
\end{equation}
and, for a trial ion-pair addition of the ion pair ($j,N_{salt} \rightarrow i,N_{salt}+1$),
\begin{equation}\label{trial_ins}
    %\alpha_{j,N_{salt} \rightarrow i,N_{salt}+1}
    \alpha(j,N_{salt}\rightarrow i,N_{salt}+1)
        = \bigg(\frac{1}{V}\bigg)^2 
       \frac{\exp(-\beta K_j)}{\mathcal{N}_k},
\end{equation}
where $\mathcal{N}_k$ is the normalization constant for the kinetic term, a function of temperature and the total number of particles in the system, including a flying ion pair.
Thus, $\mathcal{N}_k$ is identical in both $i$ and $j$ states in exchange of an ion pair at the same temperature.
All momenta here are in 3D and no momenta need to be assigned in the "vertical" direction, along which an external force will determine the vertical velocity of flying ions, and will remain the others in 3D at zero altitude.
With the enforced detailed balance condition between states $i$ and $j$,
the Metropolis acceptance probability for trial insertions, $f_{ins}$, is:
\begin{equation} \label{acc_ins}
\begin{split}
    f_{ins} 
    & = \min\bigg[
    1,\frac{acc(j,N_{salt}\rightarrow i,N_{salt}+1)}{acc(i,N_{salt}+1\rightarrow j,N_{salt})}
    \bigg], \\
\end{split}
\end{equation}
where
\begin{equation}\label{metro_ins}
\begin{split}
& \frac{acc(j,N_{salt}\rightarrow i,N_{salt}+1)}{acc(i,N_{salt}+1\rightarrow j,N_{salt})} \\ 
    & =\exp(-\beta\Delta H_{N_{salt}\rightarrow N_{salt}+1})\bigg(\frac{V}{\Lambda_s^{3}}\frac{1}{N_{salt}+1}\bigg)^2\times \\
    &\exp(\beta\mu_{salt})= \exp(-\beta\Delta M+\beta\mu_{salt}),
\end{split}
\end{equation}
%\end{widetext}
where $\Delta M$ is the nonequilibruim work for the trial insertion, $\Delta H=\Delta U + \Delta K$ the total mechanical energy difference, $\mu_{salt}$ the chemical potential, and $\Lambda_s  (=\sqrt{\Lambda_+\Lambda_-})$ the geometrical mean of thermal de Brogile wavelengths of a salt pair.
The first equality uses Equations~\ref{trial_del} - \ref{trial_ins}, and the second equality is the definition of $\Delta M$.
Accordingly, the Metropolis acceptance probability for trial deletions, $f_{des}$ is:
\begin{equation} \label{acc_des}
\begin{split}
    f_{des} 
    & = \min\bigg[
    1,\frac{acc(i,N_{salt}+1\rightarrow j,N_{salt})}{acc(j,N_{salt}\rightarrow i,N_{salt}+1)}
    \bigg], \\
\end{split}
\end{equation}
where
%\begin{widetext}
\begin{equation}\label{metro_del}
\begin{split}
    &\frac{acc(i,N_{salt}+1\rightarrow j,N_{salt})}{acc(j,N_{salt}\rightarrow i,N_{salt}+1)} \\
    &=\exp(-\beta\Delta H_{N_{salt}+1\rightarrow N_{salt}})\bigg[(N_{salt}+1)\frac{\Lambda_s^{3}}{V}\bigg]^2\times\\
    &\exp(-\beta\mu_{salt}) = \exp(+\beta\Delta M-\beta\mu_{salt}).
\end{split}
\end{equation}
%\end{widetext}
Here, the opposite sign of $\Delta M$ for the trial deletion is due to the fact that $\Delta M$ is defined for the trial insertion $N_{salt}\rightarrow N_{salt}+1$ as in Ref.~\onlinecite{belloni2019non}.
In this work, we set $\Lambda_s=\sqrt{\Lambda_+\Lambda_-}=1$ in unit length (e.g., $\Lambda_s=1$ \AA$ $ for aqueous solutions or $\Lambda_s=1\sigma$ for Lennard-Jones electrolytes). 

\textbf{Optimal choice of parameters for the maximal efficiency of H4D.}\label{subsec:eff_h4d}
The efficiency of H4D for single exchange primarily depends on a choice of $w_{max}$ and $v_f$ (or $|t_f-t_i|=|w_{max}/v_f|=\delta t\cdot N_f$), both of which determines the time spent in NEMD, where $\delta t$ is a timestep and $N_f$ is the number of integration steps during NEMD (see Figure~\ref{h4d}).
Thus, there are three free parameters for the maximal efficiency.
The optimal choice of the parameters, which is system-specific, should be found by maximizing $E_f=P_{acc}\cdot{N_f}^{-1}$, \textit{i.e.,} acheiveing a trade-off between the acceptance rate and the number of energy or force evaluations, as in any hybrid methods.
In practice, $w_{max}$ is about the size of flying particles; too small $w_{max}$ compared to the size of a flying particle may create too large initial overlap between a flying particle and other particles, and too big $w_{max}$ may require too long simulation time in NEMD at a fixed $v_f$.
For a given $w_{max}$ and $\delta t$, the optimal $v_f$ can be determined by comparing $E_f$, an efficiency metric of H4D, at several $N_f$ (thereby $v_f$).
We also stress that $\delta t$ plays a role in determining $P_{acc}$, which in NEMD is not necessarily to be as short as in equilibrium MD (\textit{e.g.}, $\delta t\approx 6$ fs was used in Ref~\onlinecite{belloni2019non}). 
A big $\delta t$ reduces $N_f$ for the same $v_f$, although it leads to a large fluctuation in $\Delta H$ along particle exchange, thereby reducing $P_{acc}$. 

As is a general idea of hybrid methods, including H4D, increasing $N_f$ (decreasing $v_f$) leads to enhanced $P_{acc}$;
$P_{acc}$ eventually saturates to a particularly value (less than unity in practice) with a large $N_f$.
Thus, there is a maximum $E_f$ at a particular $v_f$ for the maximal efficiency of H4D.
$P_{acc}$ can be estimated using the nonequilibrium work distributions and the further discussion will be given in Section~\ref{subsec:chempot_analyzing}.
We note that $E_f$ is ill-defined for the conventional MC with $N_f=0$.
In such a case, 
we compared $E^t_f$ $=P_{acc}/(N_{eq}+N_f)$, which takes into account a total simulation time of both equilibrium and nonequilibrium MD for single exchange.
We used $E^t_f$ when comparing the efficiency between H4D and the conventional MC, although $E^t_f$ could be biased towards H4D.
In the following Section, we discuss the bias techniques to obtain higher $P_{acc}$, \textit{i.e.,} enhanced efficiency of H4D, at given the other H4D parameters.

\subsection{"Tricks" for higher $P_{acc}$ in ion-pair exchange in electrolyte solutions}\label{subsection:trick}
For electrolyte systems, the major bottleneck in H4D is the exchange of ion-pair due to the strong and long-range charge-dipole and charge-charge interactions in addition to short-range overlaps.
For better efficiency (enhanced $P_{acc}$), we take advantage of the flexibility to design a protocol during NEMD, including a nonequilibrium (NE) potential energy surface, and some bias techniques.

In this section, we discuss details of the "tricks" to facilitate the exchange of an ion-pair, some of which are briefly introduced in Ref \onlinecite{belloni2019non}.
They include:
(i) fixing 3D positions of a flying-ion pair (highly recommended),
(ii) screening electrostatic interactions in both short- and long-range contributions (highly recommended),
(iii) long cut-off distance for interactions (highly recommended),
(iv) biasing distance between the two flying ions (necessary), and
(v) restricting a 3D region where particles relax, centered around a flying particle (necessary for large systems).
In addition, we implemented an early rejection scheme that is usually employed in GCMC simulations \cite{}, which allows to avoid wasting simulation times for highly unfavorable initial configurations that are rare with the bias in the flying ion distances.
The discussion about our early rejection scheme is given in the SI.

\subsubsection{\textbf{Fixing 3D position of flying ions}}
We fix the 3D position of flying ions for two reasons. 
Firstly, flying ions tend to aggregate due to strong Coulomb attraction when they are partly desolvated from bulk 3D electrolytes at non-zero altitude. 
Secondly, it is better to polarize the non-flying electrolytes in the same 3D region during NEMD, which is more likely to result in a favorable final configuration.
Here, the other ions, that are already present in the 3D electrolyte, are free to move during NEMD, although in some cases, such a dilute electrolytes, a larger acceptance rate might be achieved
with all the ions fixed in 3D space.
We note that fixing 3D positions of non-flying ions could result in better efficiency, allowing for a higher $v_f$.
The enhancement could be achieved only with a bias to choose initial positions of flying ions in order to avoid steric clashes with non-flying particles.
In our implementation, however, no such a bias was included.

\subsubsection{\textbf{Screening short- and long-range contributions to 4D electrostatic interactions}} \label{subsubsec:scr_4d_coul}
During NEMD, we need to calculate the 4D Coulomb interaction potential, $U_{Coul}(\{\vec{r}\},w)$, which depends on the altitude, $w$.
Following the standard Ewald summation method \cite{frenkel2001understanding}, we separate the short-range (sr) and long-range (lr) contributions of 4D Coulomb interactions, and introduce additional screening functions acting along the altitude:
\begin{equation}\label{4d_coul_split}
\begin{split}
&U_{Coul}(\{\vec{r}\},w) \\
&=U_{Coul,sr}(\{\vec{r}\},w)+U_{Coul,lr}(\{\vec{r}\},w) \\
& \approx U_{Coul,sr}(\{\vec{r}\},w)\cdot g^{scr,sr}(w) \\
&+U^{3D}_{Coul,lr}(\{\vec{r}\})\cdot g^{scr,lr}(w).
\end{split}
\end{equation}
The approximation includes two altitude-dependent screening functions, apart from the standard Ewald screening.
On the one hand, $g^{scr,lr}(w)$ is to approximate the exact non-analytical 4D long-range Coulomb contribution with the 3D counterpart $U^{3D}_{Coul,lr}(\{\vec{r}\})$.
On the other hand, $g^{scr,sr}(w)$ is an arbitrary function to modify 4D short-range Coulomb contribution for enhanced acceptance rate, which does a duplicate action in a sense that $U_{Coul,sr}(\{\vec{r}\},w)$ is already "screened" by the standard Ewald screening.
The modified 4D short-range part does not affect the Metropolis acceptance rule for ion-pair exchange as long as NEMD is evolved with a deterministic propagator and a time-dependent altitude protocol is pre-determined.

The idea of the additional screening short-range Coulomb contribution along the altitude is similar to the potential scaling method \cite{inagaki2022hybrid} or the replica-exchange with solute tempering \cite{wang2011replica}.
$g^{scr,sr}$ flattens only the forces associated with flying ions, and thereby acts "locally".
The local flattening is expected to facilitate a barrier-crossing event along the associated degrees of freedom, \textit{e.g.}, ion-pair exchange in our case.

Among various choices, we chose an exponential function for both $g^{scr,lr}(w)$ and $g^{scr,sr}(w)$.
\begin{equation}\label{eq:scr_4d_coul_lr}
g^{scr,lr}(w)=\exp(-\kappa^{lr}_w w),
\end{equation}
and
\begin{equation}\label{eq:scr_4d_coul_sr}
g_{ij}^{scr,sr}(w)=\exp(-\kappa^{sr}_w w)\cdot f_{ij},
\end{equation}
where $f_{ij}$ is unity only when one of $i$ and $j$ particles is a flying one, and zero otherwise.
$\kappa^{lr}_w$ and $\kappa^{sr}_w$ are the screening parameters for the short- and long-range contributions along the vertical axis, respectively.
Thus, the short-range screening function only acts on the Coulomb interactions between flying and non-flying one, while $g^{scr,lr}(w)$ screens all the long-range contributions of Coulomb interactions uniformly. While it would in principle be possible to only screen the interactions involving the flying ions also for the long-range contribution, in practice our choice is easier to implement. We found that a screening function that uniformly screens the short-range Coulomb interactions is not beneficial in enhancing efficiency.
In the case of the exchange of multiple particles with different time-dependent altitudes, we use an average altitude ($\bar{w}$) among flying particles, \textit{i.e.}, $g^{scr,lr}=g^{scr,lr}(\bar{w})$, and $g^{scr,sr}=g^{scr,sr}(\bar{w})$.
We also note that another choice of $g^{scr,lr}(w)$ could be a Gaussian function (as in Ref~\onlinecite{belloni2019non} even though no explicit expression was given), which naturally appears in approximating the exact 4D Coulomb interactions.

We want to stress again that the choice of the screening functions does depend on the system of interest.
For example, we found $g^{scr,lr}(w)=1$ ($\kappa^{lr}_w=0$) gives better results for LJ electrolytes. 
Another choice might be Wolf potential which needs no long-range contribution of Coulomb interactions.
We found no benefits in the efficiency with the truncated and shifted Wolf potential for the example systems considered in this work. 

\subsubsection{\textbf{Cut-off distance for the short-range part of 4D Coulomb interactions}} 

In calculating the short-range part of Coulomb interactions both in 3D during equilibrium MD or 4D during NEMD, one sets a cut-off distance and in LAMMPS, the Ewald screening parameter is chosen for the particular cut-off distance so as to achieve a given relative error in forces ($10^{-4}$ in this work). 
We found markedly enhanced efficiency in ion-pair exchange when increasing the cut-off distance for $U_{Coul,sr}(\{\vec{r}\},w)$ with given $w$-dependent screening parameters ($\kappa^{lr}_w$ and $\kappa^{sr}_w$).
If one cuts $U_{Coul,sr}(\{\vec{r}\},w)$ at a short distance, the efficiency drops significantly even with other tricks discussed above. Thus, a long cut-off distance is preferred during NEMD. For example, for aqueous electrolytes, we use a cut-off of 14~\AA$ $ is chosen during NEMD, instead of 9~\AA$ $ during equilibrium MD. However, we did not explore the effect of the cut-off distance on the efficiency of H4D with a fixed Ewald screening parameter, as we try to minimize the changes in the LAMMPS source code. We also found that increasing the cut-off distance for LJ interactions is much less significant, and we used the same value for NEMD and equilibrium MD.

\subsubsection{\textbf{Biasing the distance between flying ions in favor of short ones}}\label{sec:bias}
For an enhanced efficiency, we biased the distance between flying ions in favor of short distances as was briefly mentioned in the SI of Ref~\onlinecite{belloni2019non}.
According to the Stillinger-Lovett sum rules \cite{stillinger1968ion,stillinger1968general}, the charge neutrality applies even at short distances on the order of the Debye screening length. 
Thus, it is unfavorable to investigate larger distances between flying cation and anion.
Furthermore, a small dipole of a flying-ion pair to exchange could limit its perturbation to an electrolyte system.
Since the 3D position of flying ions is fixed during NEMD, it is entirely determined by their initialization.
In case of trial insertion,  instead of randomly placing both flying ions, the choice of an initial position ($\vec{r}_a=(x_a,y_a,z_a)$) for the flying anion depends on a randomly chosen initial position ($\vec{r}_c=(x_c,y_c,z_c)$) of the flying cation.
For simplicity, the same $w_{max}$ is assumed for both flying ions in this discussion.

Bias for trial insertion and deletion must be chosen consistently to satisfy the detailed balance.
In the case of a trial insertion, for the position of a flying anion, knowing that of a flying cation,
\begin{equation}
B^{ins}(\vec{r}_a,\vec{r}_c)=b^{ins}(x_a|x_c)b^{ins}(y_a|y_c)b^{ins}(z_a|z_c)\cdot V,
\end{equation}
where $V=L_xL_yL_z$ is the volume of the simulation box.
Without such a bias, $b^{ins}=1/L_i$ for each direction $i  (\in \{x,y,z\}$), and thereby $B^{ins}=1$.
In the case of a trial deletion, choosing one of the $N_{salt}+1$ anions, knowing $\vec{r}_c$ of a flying cation,
\begin{equation}
B^{del}(\vec{r}_a,\vec{r}_c)=\frac{B^{ins}(\vec{r}_a,\vec{r}_c)}{\sum_{n=1}^{N_{salt}+1}B^{ins}(\vec{r}_{a,n},\vec{r}_c)}\cdot(N_{salt}+1).
\end{equation}
In this case, we use the same functional form for the bias in both insertion and deletion trial moves, but one can use a different functional form for each bias.
Again, $b^{ins}=1/L_i$, and $B^{del}=1$ in the absence of such a bias.

There are several choices for such a biasing function (e.g., a Gaussian distribution with a 3D distance between flying ions). In this work, we chose a bimodal distribution function for $b^{ins}$.
This bimodal distribution helps to generate a pair of flying ions that are close to each other yet without high overlap.
\begin{widetext}
\begin{equation}\label{eq:bimodal}
    b^{ins}(x_a|x_c)=
    \sqrt{\frac{\alpha_b}{\pi}}
    \frac{\exp(-\alpha_b(x_{ac}-x_b)^2)+\exp(-\alpha_b(x_{ac}+x_b)^2)}
    {\text{erf}[\sqrt{\alpha_b}(x_b+L_x/2)]-\text{erf}[\sqrt{\alpha_b}(x_b-L_x/2)]}, 
\end{equation}
\end{widetext}
where $x_{ac}=x_a-x_c$. 
$x_b$ determines the mean separation of flying ions, and $\alpha_b$ determines the sharpness of the biasing function.
Again, $b^{ins}$ is normalized such that $\int_{-L_x/2}^{L_x/2}b^{ins}(x_a|x_c)dx_a=1$.

With the bias, the Metropolis acceptance rule should be modified accordingly, similar to the Rosenbluth factor in biased MC techniques \cite{rosenbluth1955monte,frenkel2001understanding}.
The acceptance probability for trial insertions ($N_{salt} \rightarrow N_{salt}+1$) becomes:
\begin{widetext}
\begin{equation}\label{metro_ins_bias}
\begin{split}
    f^{bias}_{ins}(\vec{r}_a,\vec{r}_c)
    & =\min\bigg[1,
    \exp(-\beta\Delta H_{N_{salt}\rightarrow N_{salt}+1})\exp(\beta\mu_{salt})
    %\bigg(\frac{1}{VG^b(\vec{r}_a|\vec{r}_c)P_{sel}}\bigg)
    \bigg(\frac{V}{\Lambda_s^{3}}\frac{1}{N_{salt}+1}\bigg)^2
    \frac{B^{del}(\vec{r}_a|\vec{r}_c)}{B^{ins}(\vec{r}_a|\vec{r}_c)}
    \bigg],
\end{split}
\end{equation}
\end{widetext}
while the acceptance probability for trial deletions ($N_{salt}+1 \rightarrow N_{salt}$), which also depends on the the selection of trial positions of flying ions, becomes:
\begin{widetext}
\begin{equation}\label{metro_del_bias}
\begin{split}
    f^{bias}_{des}(\vec{r}_a,\vec{r}_c)
    & = \min\bigg[1,
    \exp(-\beta\Delta H_{N_{salt}+1\rightarrow N_{salt}})\exp(-\beta\mu_{salt})
    \bigg((N_{salt}+1)\frac{\Lambda_s^{3}}{V}\bigg)^2
    \frac{B^{ins}(\vec{r}_a|\vec{r}_c)}{B^{del}(\vec{r}_a|\vec{r}_c)}
    \bigg].
\end{split}
\end{equation}
\end{widetext}

\subsubsection{\textbf{Restricting a 3D region to relax}} 
The efficiency of a hybrid method, which employs global relaxation, is known to be lower with a bigger system size due to the larger fluctuations in the total energy \cite{frenkel2001understanding}.
In principle, it can therefore be useful to consider an algorithm exploiting only a (quasi-)local relaxation, instead of a global one.
One possibility for particle exchange is to restrict the relaxation to a 3D region centered around a flying particle, as suggested in Ref~\onlinecite{belloni2019non}.
In this case, another factor, $f_{sel}(\text{"old"}\rightarrow\text{"new"})$, should be taken into account for the Metropolis acceptance rule to maintain the detailed balance:
\begin{equation}
    f_{sel}(\text{"old"}\rightarrow\text{"new"})=\frac
    {\prod_{i\in \mathcal{I}_{sel}}p_{sel}(r_{i,\text{new}})}
    {\prod_{i\in \mathcal{I}_{sel}}p_{sel}(r_{i,\text{old}})},
\end{equation}
where $\mathcal{I}_{sel}$ is an index set of particles in the region to relax, and $p_{sel}$ is an activation function for the selection process to determine whether or not the $i^{th}$ particle at a distance $r_i$ from the flying particle is allowed to relax.
Here, "old" and "new" refer to a configuration before and after NEMD, including flying particles, respectively.

One example is a spherical region centered around a single flying particle. 
In this case, one can use a hyperbolic tangent function as an activation function, which depends only on the distance from the center of the region:
\begin{equation}\label{eq:soft_act}
   p_{sel}(r_{i})=\frac{1}{2}\bigg(
   1-\tanh\bigg(\frac{r_i-l_{sel}}{w_{sel}}\bigg)
   \bigg),
\end{equation}
where $l_{sel}$ determines the size of the spherical region, and $w_{sel}$ determines the stiffness of the activation function at the boundary.
Due to the reversibility, a soft boundary should be used with a non-zero $w_{sel}$.
Otherwise, a proposed trial move using H4D should be automatically rejected as soon as one of non-flying particles enters into or leaves the selected region.
Then, the index set should be $\mathcal{I}_{sel}=\{i|\xi\le p_{sel}(r_{i,\text{old}})\}$ with a random number, $\xi$, drawn for each particle from a uniform interval between 0 and 1.
This index set should be determined using an "old" configuration before NEMD.

We note that $f_{sel}$ is usually less than unity, which means that it will reduce the acceptance rate. To minimize this downside, one can introduce an external force acting on the particles inside the relaxing region to keep them inside the region, but we did not consider this in the present work.
In principle, this method should work for all systems, and benefit big systems by reducing the fluctuations in total energy as well as the number of particles to be considered in the propagation of the NEMD trajectory (the other ones, in majority, being fixed). However, we found that there is no additional benefit from this method for both electrolytes in this work, and will not discuss the results here.

\subsection{Implementation in LAMMPS}
We implemented the H4D method described above in LAMMPS \cite{plimpton1995fast}, an open-source molecular dynamics simulation package for a general purpose, allowing for massively parallel calculations.
Grand-canonical Monte Carlo simulations and a hybrid method with constant particle number \cite{guo2018hybrid} are already available in LAMMPS. 
And yet, a script for GCMD simulations in LAMMPS is not available to our knowledge, particularly with the advanced methods of Section~\ref{subsection:trick}.

Our implementation (See Data Availability), done with a LAMMPS version released on Oct. 27, 2021, includes several c++ files for NEMD and a python wrapper to control GC MD simulations.
On the one hand, the c++ files for NEMD include the time-dependent altitude schedule that determines a nonequilibrium potentail energy surface (NE PES), as discussed above. 
They have only a few modifications on existing c++ files, such as a "pair$\_$lj$\_$cut$\_$coul$\_$long.cpp" file, including calculation of the 4D distance with the altitude between flying and non-flying particles, and NE PES with the screening of 4D electrostatics to both short- and long-range contributions as discussed above.
Thus, a different c++ file is required to run simulations with a different NE PES.
On the other hand, the Python wrapper does several jobs, including all the calculations related to the Monte Carlo step in particle exchange.
Our implementation is flexible enough to apply to various systems, including confined electrolytes.

\subsection{Model electrolytes}\label{subsec:model}
The H4D method can be used for all systems that require the exchange of neutral or charged particles, such as Lennard-Jones (LJ) fluids, pure water, or electrolytes \cite{belloni2019non}.
In the present work, we illustrate the challenges related to ion-pair exchange that involves long-range Coulomb interactions in electrolytes. 
In this Section, we introduce the two model electrolytes considered and the computer simulation details.

\subsubsection{Lennard-Jones electrolytes}
We first consider LJ electrolytes that consist of neutral LJ solvent particles and ion particles, all of which are of the same size and of same mass ($m$) \cite{joly2006liquid}.
All the LJ interactions ($U_{LJ}$) were truncated and shifted at a cut-off distance $r^*_c=r_c/\sigma=2.5$.
\begin{equation}
\begin{split}
U_{LJ}(r)=&4\epsilon\bigg[\bigg(\frac{\sigma}{r}\bigg)^{12}-
\bigg(\frac{\sigma}{r}\bigg)^{6}\\
&-\bigg(\frac{\sigma}{r_c}\bigg)^{12}+\bigg(\frac{\sigma}{r_c}\bigg)^{6}\bigg],
\end{split}
\end{equation}
assuming the same LJ energy $\epsilon$, and diameter $\sigma$ for interactions between all types of particles.
Here, the asterisk represents a quantity in reduced LJ unit.
The Coulomb interaction ($U_C$) between LJ ions is:
\begin{equation}
U_{C}(r)=\frac{1}{4\pi\epsilon_0\epsilon_s}\frac{q_iq_j}{r}=\frac{\epsilon}{\epsilon_s}\frac{q^*_iq^*_j}{r^*},
\end{equation}
where $\epsilon_0$ is the vacuum permittivity, and $\epsilon_s$ is the uniform background dielectric constant, either 1 or 0.2 for a high or low dielectric solvent.
LJ ions carry either $q^*_i=q_i/\sqrt{4\pi\epsilon_0\sigma\epsilon}=+1$ or -1.
As in Equation~\ref{4d_coul_split}, $U_C(r)$ is split into two contributions, following the standard Ewald summation technique: The short-range contribution is cut at $r^*=3.5$, and the long-range contribution is calculated using the particle-particle and particle-mesh (PPPM) method with the Ewald screening parameter chosen for a fixed cut-off distance in order to achieve a given relative error in forces ($10^{-4}$ in this work). 
Configuration sampling to calculate the chemical potentials was done in the $N_{solv}N_{salt}p^*T^*$ ensemble at $p^*=p\sigma^3/\epsilon=1$ and $T^*=Tk_B/\epsilon=1$, which corresponds to a liquid phase \cite{scalfi2021gibbs}.
$N_{solv}$ was fixed to 5000, and $N_{salt}$ varies according to the solution molality, unless otherwise noted.

During equilibrium MD, the equations of motion were integrated using the velocity Verlet integrator with a timestep $\delta t^*=\delta t\sqrt{\epsilon(m\sigma^2)^{-1}}=0.005$.
In NEMD, the velocity Verlet integrator was also used but considering timesteps $\delta t^*=0.002$, 0.005, 0.01, and 0.02.
We found that $\delta t^*=$0.02 during NEMD results in negligible acceptance rate of particle exchange due to large total energy fluctuations.
In this work, the maximum altitude for H4D, $w^*_{max}$, was set to unity for both solvent and ion-pair exchange in LJ electrolytes.

In regards to treating the 4D electrostatics during NEMD, we found two things in case of the LJ electrolytes.
Firstly, as expected, the PPPM method is much faster than Ewald summation to calculate a long-range contribution of Coulomb interactions at high solution molalities with more than 10,000 ions.
Secondly, approximating 4D PPPM to 3D PPPM ($U^{4D}_{Coul,lr}(\{\vec{r}\},w)=U^{3D}_{Coul,lr}(\{\vec{r}\})$ with $\kappa^{lr}_w=0$; see Equation~\ref{4d_coul_split}) with no non-zero altitude contribution to the long-range part considered) results in higher acceptance rate for both solvent and ion-pair exchange, while the short-range Coulomb contribution was calculated in 4D.

\subsubsection{Aqueous NaCl electrolytes}
As a more realistic system, we also investigated aqueous NaCl electrolytes.
We chose a force field, combining the SPC/E water model \cite{berendsen1987missing} with the Joung-Cheatham one for the ions \cite{mouvcka2013molecular,mester2015mean}, which has been widely used to study salt solubility and mean ion activity coefficients.
All the LJ interactions were truncated and shifted at $r=9$ \AA$ $.
Coulomb interactions were cut at $r=9$ \AA$ $, and its long-range contribution was calculated using PPPM with the Ewald screening parameter chosen for a fixed cut-off distance in order to achieve a given relative error in forces ($10^{-4}$ in this work).
The Lorentz-Berthelot mixing rule was applied for the cross-interaction parameters.
Configuration sampling to calculate the chemical potentials was done in $N_{\text{water}}N_{\text{NaCl}}$$pT$ ensemble at $p=1$ atm and $T=298.15$ K for a wide range of solution molarity from dilute to concentrated regimes, unless otherwise noted. $N_{\text{water}}$ was fixed to 500, and $N_{\text{NaCl}}$ varied according to the solution molality.

During equilibrium MD, the equations of motion were integrated using the velocity Verlet integrator with a timestep $\delta t=2$ fs, and the SHAKE algorithm \cite{andersen1983rattle} was used to treat water molecules as rigid.
In NEMD, the velocity Verlet integrator was used only for ions, but a quaternion-based integrator \cite{miller2002symplectic} was used for rigid water molecules instead of SHAKE. A timestep of $\delta t=4$ fs was used in NEMD. Unlike in Ref.~\onlinecite{belloni2019non}, we did not consider identical masses for the O and H atoms in water molecules for the NEMD step.
%In aqueous electrolytes, there were no numerical issues with this $\delta t$ with little more frequent neighbor list check.
In this work, $w_{max}$ was set to be 3 \AA $ $ for both water and ion-pair exchange in aqueous electrolytes, as was suggested in Ref~\onlinecite{belloni2019non}.

\subsection{Estimating $P_{acc}$ from the nonequilibrium work distributions}\label{subsec:chempot_analyzing}

Following Equations~\ref{metro_ins} and~\ref{metro_del}, $P_{acc}$ can be estimated using the nonequilibrium work distributions:
\begin{widetext}
\begin{equation}\label{pacc_theory}
\begin{split}
P_{acc}& = 
\begin{cases}
\bigg\langle \min[1,\exp(-\beta\Delta M+\beta\mu]\bigg\rangle \text{ for trial insertion moves} \\
\bigg\langle \min[1,\exp(\beta\Delta M-\beta\mu)]\bigg\rangle \text{ for trial deletion moves}
\end{cases}\\
& =
\begin{cases}
    R^{ov}_{ins} + \int_{\mu}^{\infty} P_{ins}(\Delta M)\exp{(-\beta\Delta M+\beta\mu)} d\Delta M  \text{ for trial insertion moves}\\
    R^{ov}_{del} + \int_{-\infty}^{\mu} P_{del}(\Delta M)\exp{(\beta\Delta M-\beta\mu)} d\Delta M \text{ for trial deletion moves.}
\end{cases}
\end{split}
\end{equation}
\end{widetext}
Here, $R^{ov}_{ins}=\int_{\mu_s}^{\infty} P_{ins}(\Delta M) d\Delta M$ and $R_{ov,del}=\int_{-\infty}^{\mu_s} P_{del}(\Delta M) d\Delta M$.
$\langle\cdots\rangle$ is the ensemble average at fixed particle numbers of both solvent and salt, so no trial moves must be accepted in sampling $P_{ins}(\Delta M)$ and $P_{del}(\Delta M)$.
In the H4D, both the distributions get closer to each other with smaller variances with a slower $v_f$ (a larger $N_f$), resulting in a higher $P_{acc}$.
Once the work distributions are Gaussian (with sufficiently small $\delta t$ and $v_f$), $P_{acc}$ can be estimated analytically with the well-known equation \cite{mehlig1992hybrid}: $P_{acc}=\text{erfc}(0.5\sqrt{\beta\langle\Delta H\rangle})$, where $\text{erfc}(\cdot)$ is a complementary error function.
However, in general, the work distributions are not Gaussian; for example, at a fast $v_f$ the distribution for trial insertions usually exhibits a fat tail due to an unfavorable trial exchange.
Further, the biased distances (Section~\ref{sec:bias}) between flying ions could result in highly non-Gaussian work distributions, as will be discussed in the Result Section.
In this work, the non-Gaussianity was quantified as follows:
\begin{equation}
    \alpha_2=\frac{1}{3}\frac{\langle(\beta\delta M)^4\rangle}{\langle(\beta\delta M)^2\rangle^2}-1.
\end{equation}
A positive (negative) $\alpha_2$ means a broader (narrower) distribution than the estimated Gaussian with the mean and variance.
Furthermore, similarly to the Widom method \cite{widom1963some}, chemical potentials can be computed using the nonequilibrium work distributions in H4D via Crooks' fluctuation theorem \cite{crooks1999entropy}, as is discussed in the SI.

\section{Results and discussion}\label{results}
In this section, we discuss the results obtained using the H4D method for the LJ and aqueous NaCl electrolytes, including the calculation of chemical potentials, and GCMD simulations.

\subsection{LJ electrolytes}
\subsubsection{Chemical potential calculation using work distributions}\label{subsubsec:lj_chem_calc}
\begin{figure*}[htbp]
    \includegraphics [width=5in] {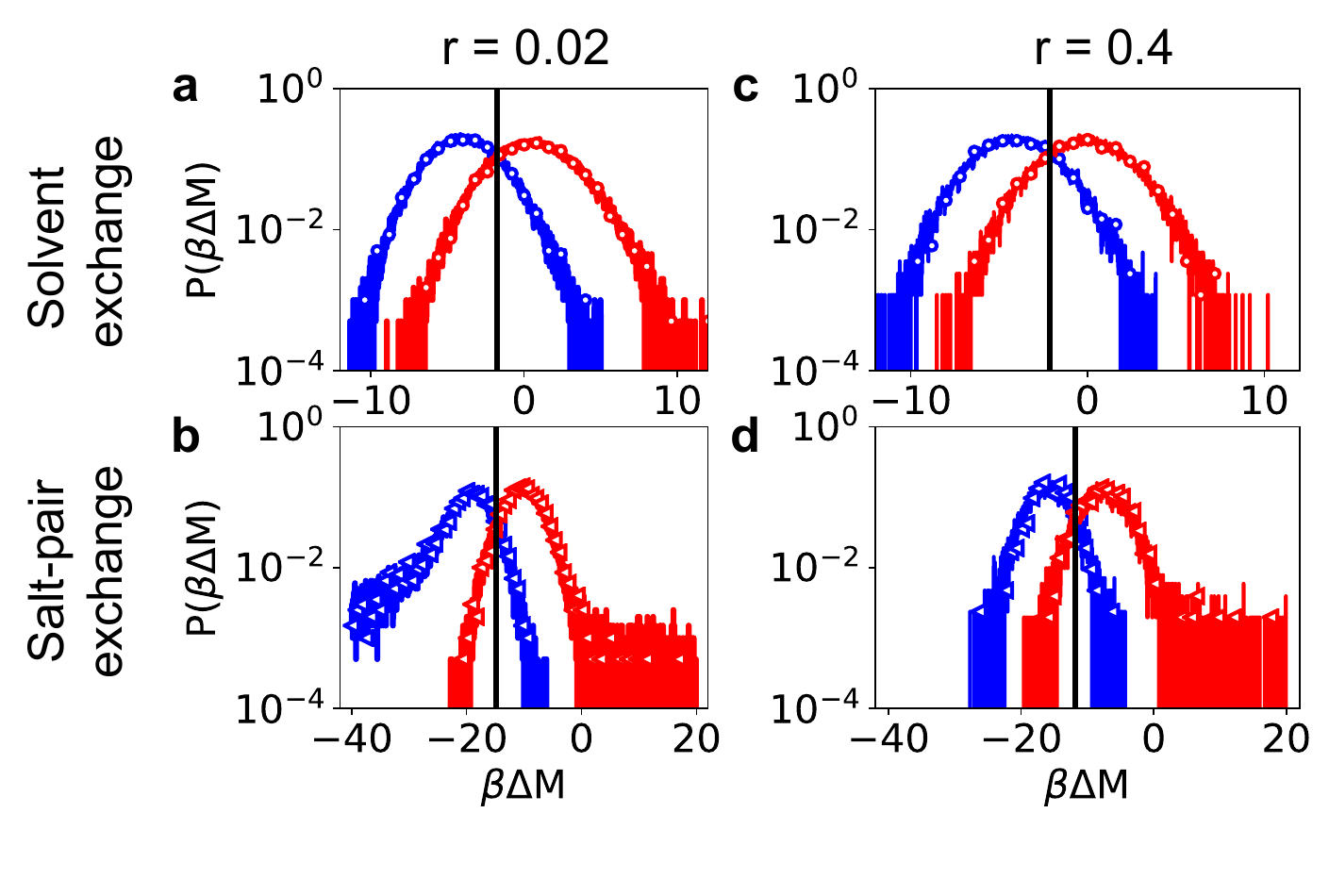}    
    \label{spm_e02_work_conc}
    \caption{Nonequilibrium work distributions of solvent (a,c) and salt-pair (b,d) exchange in dilute (left column) and concentrated (right column) LJ electrolytes ($\epsilon_s=0.2$) for fixed number ratios between the numbers of ion pairs and solvent particles: $r=N_{\text{salt}}/N_{\text{solv}}$.
Blue is for trial deletions and red for trial insertions.
Black vertical lines represent the calculated chemical potential ($\beta^*\mu^*$).
A bimodal biasing function (Eq.~\ref{eq:bimodal}) was used for the salt-pair exchange with the parameters of $\alpha^*_b=5$ and $x^*_b=0.8$.
See Table~\ref{tab:spm_e02_work_conc} for the parameters used.
}
\end{figure*}
\begin{table}[ht!]
\label{tab:spm_e02_work_conc}
    \centering
    \begin{tabular}{ |c|c|c|c|c||c| } 
    \hline
    $r$ & Species & $N_f$ & $\delta t^*$ & $v^*_f$ & $\beta^*\mu^*$ \\ 
    \hline
    \hline
    0.02 & Solvent & 500 & 0.01 & 0.2 & -1.754 ($\pm$0.007)  \\
    0.02 & Salt pair & 1000 & 0.01 & 0.1 & -14.81 ($\pm$0.01) \\
    \hline
    0.4 & Solvent & 1000 & 0.01 & 0.1 & -2.12 ($\pm$0.01)  \\
    0.4 & Salt pair & 2000 & 0.01 & 0.05 & -11.78 ($\pm$0.01) \\
    \hline
    \end{tabular}
    \caption{Parameters used in sampling of the nonequilibrium work distributions in Figure~\ref{spm_e02_work_conc}, and computed chemical potentials $\mu^*$ of the solvent and the salt pair at two number ratios between the numbers of ion pairs and solvent particles ($r=N_{\text{salt}}/N_{\text{solv}}$).
    In all cases, $N_{eq}=1,000$ and $w^*_{max}=1$.
}
\end{table}
    
Figure~\ref{spm_e02_work_conc} displays the nonequilibrium work distributions of trial (but never accepted) insertions and deletions using H4D of either a solvent or a salt pair in dilute and concentrated LJ electrolytes ($r=N_{\text{salt}}/N_{\text{solv}}=0.02$, and $0.4$, respectively) as described in Section~\ref{subsec:model}. The trial exchanges were performed in the $N_{\text{solv}}N_{\text{salt}}pT$ ensemble at a particular vertical velocity ($v^*_f$), and none of them were accepted after calculating the work distributions.
For each species, the work distributions $P_{ins}$ and $P_{del}$ intersect for $\Delta M^*=\mu^*$, following the CFT \cite{crooks1999entropy}, and $\mu^*$ was calculated using the BAR method as described in Section~\ref{subsubsec:cft_bar}.
The large overlap between $P_{ins}$ and $P_{del}$, as shown in Figure~\ref{spm_e02_work_conc}, helps to reduce the numerical error in estimating $\mu^*$ (we note that the chemical potentials were computed without fluctuations in composition so they may suffer from the finite-size effects)~\cite{belloni2018finite}. To do so, the slow enough $v^*_f$ was chosen, and a bimodal biasing function (Eq.~\ref{eq:bimodal}) was introduced.
The bias effect and the skewed work distributions will be discussed in detail in Section~\ref{subsection:bias_nacl_pair} in the case of NaCl-pair exchange in aqueous electrolytes.
No excluded volume for the early rejection (Section~\ref{subsection:early_rejection}) was considered in calculating the work distributions.

In general, both solvent and salt-pair exchanges appeared to be relatively facile in the LJ electrolytes than in the aqueous electrolytes, requiring smaller $N_f$ in NEMD.
That is because our model LJ solvents carry no dipole; instead, a background static dielectric constant was introduced.
In other words, there are no dipole-dipole or dipole-charge interactions, although charge-charge interactions are present.
Thus, there is no frequency-dependent solvent dielectric polarization that could slowly relax toward equilibrium during NEMD.

The solvent and salt pair's chemical potentials in a binary solution are not independent.
In our model LJ electrolytes, the calculated chemical potentials were found to satisfy the Gibbs-Duhem relationship, exhibiting thermodynamic consistency (See Section~\ref{sec:SI_gde} in SI).

\subsubsection{Grand-canonical MD simulations}
\begin {figure}[htbp]
\includegraphics [width=3in] {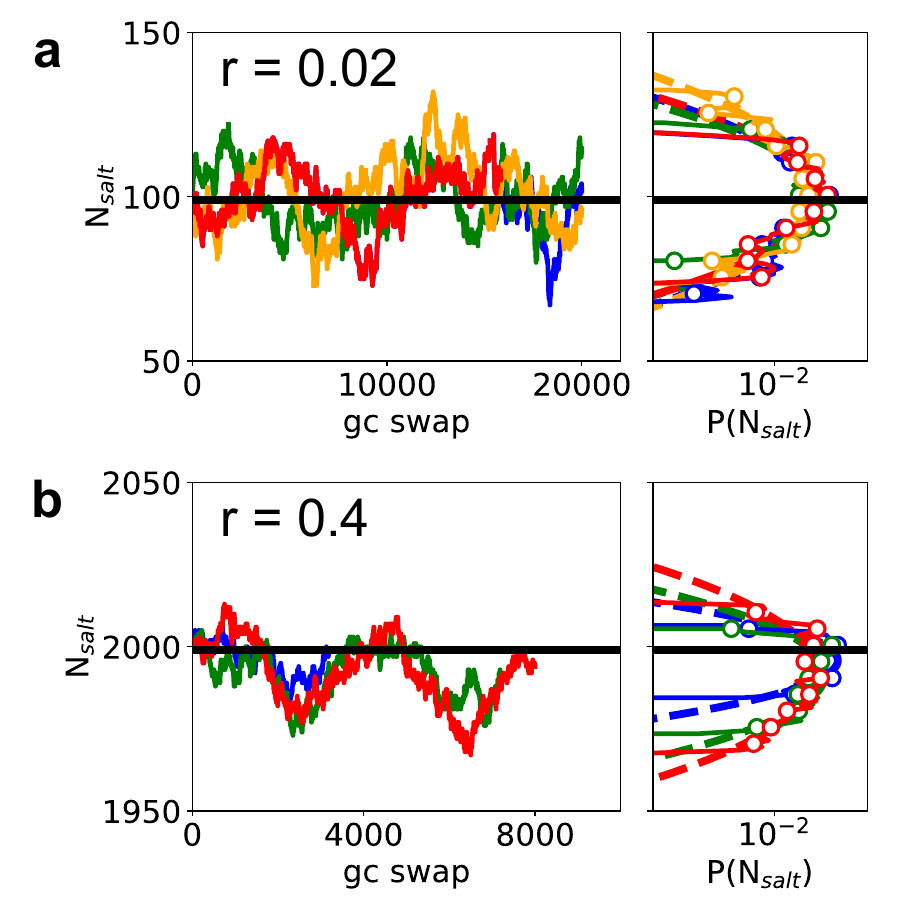}
\caption{Salt-pair number fluctuations (left) and its distributions (right) in model LJ electrolytes at the same two molar ratios $r$ in Figure~\ref{spm_e02_work_conc} during GCMD simulations.
Different colors represent different independent MD runs.
Simulations were performed in the $N_{\text{solv}}\mu^*_{\text{salt}}$$p^*T^*$ ensemble using chemical potentials determined in Section~\ref{subsubsec:lj_chem_calc}.
The black horizontal lines in both panels indicate the number of salt pairs in calculating $\beta^*\mu^*_{\text{salt}}$ (Fig.~\ref{tab:spm_e02_work_conc}).
A bimodal biasing function is applied with $\alpha^*_b=5$ and $x^*_b=0.8$ (Equation~\ref{eq:bimodal}).
A small excluded volume is also applied: $V^*_{ex}=0.027$ for early rejection (Section~\ref{subsection:early_rejection}).
Dotted lines on the right panel represent Gaussian distributions from each MD run with the measured average and variance.}
\label{spm_conc_gcmc_fluc}
\end{figure}
With the calculated chemical potential, GC MD simulations were employed with H4D.
There are three different ensembles to describe bulk binary electrolytes in open environment: $\mu^*_{\text{solv}}N_{\text{salt}}$$p^*T^*$, $N_{\text{solv}}\mu^*_{\text{salt}}$$p^*T^*$, and $\mu^*_{\text{solv}}\mu^*_{\text{salt}}$$V^*T^*$ ensembles; 
the choice of the ensemble depends on applications of interest, or the sampling efficiency.
Figure~\ref{spm_conc_gcmc_fluc} shows the number fluctuations of salt pairs in the LJ electrolytes during GCMD simulations ($N_{\text{solv}}\mu^*_{\text{salt}}$$p^*T^*$ ensemble) using the calculated chemical potential at each composition as in Figure~\ref{spm_e02_work_conc}.
In each run, the number of salt pairs fluctuates around its average, exhibiting excellent convergence: $\langle N_{salt}\rangle=100.0\pm 0.7$ and $\sqrt{\langle (\delta N_{salt})^2\rangle}=9.7\pm 0.5$ at $r=0.02$, and $\langle N_{salt}\rangle=1993\pm 1$ and $\sqrt{\langle (\delta N_{salt})^2\rangle}=8\pm 1$ at $r=0.4$.
In the concentrated electrolytes, the difference in $\langle N_{salt}\rangle$ from the reference value (2000) is only less than 0.5$\%$.

We found the H4D can achieve the several orders of magnitude enhancement in the efficiency of GCMD in the LJ electrolytes: $P_{acc}=0.13$ using the H4D with $v^*_f=0.1$, while $P_{acc}=1.7\cdot10^{-6}$ using the conventional MC ($v^*_f=\infty$), resulting in the about $3.8\cdot10^{4}$ times enhanced $E^t_f$ $(=P_{acc}/(N_{eq}+N_f))$.
The efficiency of the H4D, as expected, depends on the salt concentration: $P_{acc}=0.13$ at $r=0.02$ with $v^*_f=0.1$, and $P_{acc}=0.15$ at $r=0.4$ with $v^*_f=0.05$.
That is, the salt-pair exchange is twice efficient at $r=0.02$ than $0.4$.
As also expected, the salt-pair exchange was found less efficient than the solvent exchange; at both concentrations, solvent exchange in the $\mu^*_{\text{solv}}N_{\text{salt}}p^*T^*$ ensemble is about five-fold efficient than salt-pair exchange with $\sim 2.5$ times higher $P_{acc}$ and twice faster $v^*_f$.

We also found the effect of $\delta t^*$ on the efficiency of the H4D at a given $N_f$.
In the case of solvent exchange, despite the monotonic decrease in $v_f$ with increasing $\delta t^*$ (0.002, 0.005, 0.01, and 0.02) at a fixed $N_f=1000$, $\delta t^*=0.005$ appears to achieve the highest $P_{acc}$, and thereby $E_f$.
This indicates that too large $\delta t^*$ leads to dramatic time-discretizing error, significantly decreasing $P_{acc}$; $P_{acc}\le0.001$ with $\delta t^*=0.02$.

\subsection{Aqueous NaCl electrolytes}
We now turn to the results of optimal parameter search via $P_{acc}$ estimation and GCMD simulations for aqueous NaCl electrolytes.
The optimal parameter search for the aqueous electrolytes is much crucial as their $P_{acc}$ is much lower than the one for the LJ electrolytes, even with a bigger $N_f$ for both water and NaCl-pair exchanges.
As in the LJ electrolytes, the optimal parameters of NEMD should depend on the solution molality, which further complicates the parameter optimization.
NaCl-pair exchange, the most time-consuming step, was found to require about 40 ps long NEMD with $\delta t=4$ fs to reach $\sim 1\%$ acceptance rate at 1 $m$ NaCl concentration, with all the bias techniques.

\subsubsection{Water exchange}
\textbf{Effect of the vertical velocity.}
Water exchange needs particular care in preparing the initial moment for trial insertions and an integrator as described in Section~\ref{methods}.
We found no statistical differences in the nonequilibrium work distributions sampled with either a quaternion-based integrator \cite{miller2002symplectic} or SHAKE \cite{andersen1983rattle} during NEMD, and all the results here were obtained with the former.
In this section, we discuss the effect of $v_f$ on $P_{acc}$ of water exchange in aqueous solutions at two different molalities with $N_{water}=500$ and $N_{salt}=4$ and 36 for molalities of 0.44 and 4 $m$, respectively.

\begin {figure}[htbp]
\includegraphics [width=3.2in] {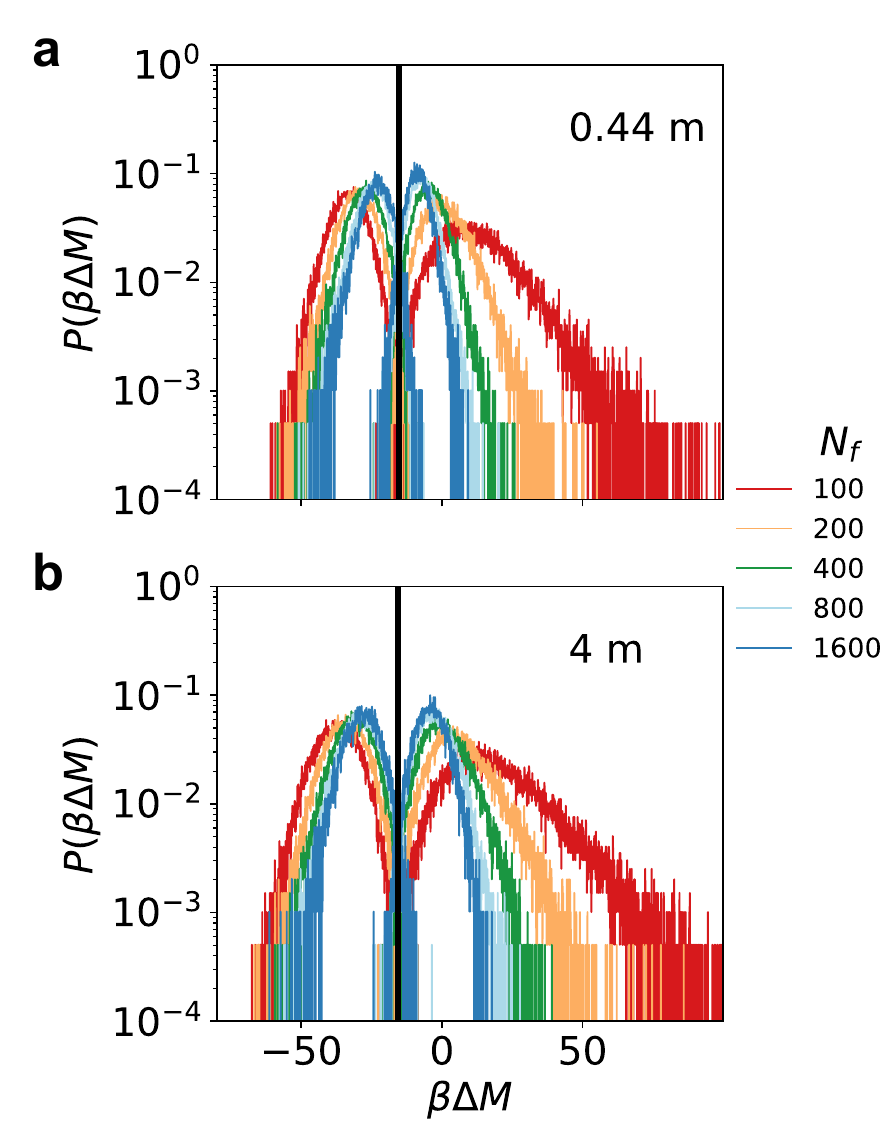}
\caption{Work distributions of water exchange at different vertical velocities in aqueous NaCl electrolytes at two molalities (a) 0.44 and  (b) 4 $m$.
$N_f$ is the number of integration steps in NEMD, so $v_f=w_{max}/(N_f\delta t)$.
Other parameters are the same: $\delta t=4$ fs, $w_{max}=3$ \AA, $N_{eq}=1000$, $\kappa^{lr}_w=1 $ \AA$^{-1}$, and $\kappa^{sr}_w=0$.
The black vertical lines indicate $\beta\mu_{\text{water}}$ \cite{mester2015mean} at each solution molality.}
\label{water_vertical}
\end{figure}
Figure~\ref{water_vertical} displays the nonequilibrium work distributions for water exchange at different vertical velocities ($v_f=\sim$0.47 - 7.5 \AA/ps) corresponding to different $N_f$ at two solution molalities.
At each molality, the distributions on the left (right) are for the water deletion (insertion) trial moves.
As is expected, the work distributions exhibit two features.
Firstly, at all $v_f$'s, the distributions of trial insertions and deletions intersect for $\Delta M=\mu_{\text{water}}$, confirming that the calculated chemical potential of water ($\mu_{\text{water}}$) is independent of $v_f$, as it should.
The resulting values ($\beta\mu_{water}=-15.39\pm0.03$ and $-15.51\pm0.05$ for molalities of 0.44 and 0.4, respectively, with $N_f=800$) are in good agreement with those reported in Ref.~\onlinecite{mester2015mean} using the method of thermodynamic integration for the same systems \cite{chempotconv}.
Secondly, $P_{acc}$ is increased with decreasing $v_f$ (increasing $N_f$) as the work distributions for both trial moves become narrower with smaller variance $\langle(\delta\beta\Delta M)^2\rangle$ and and its average $\langle\beta\Delta M\rangle$ gets closer to $\mu_{\text{water}}$ at both molalities (Table~\ref{table:water_vertical}).
It is obvious that at fast $v_f$ the Gaussian approximation for the work distributions should not work, particularly for the trial insertions, while all the distributions at small $v_f$ are almost Gaussian with fairly small $\alpha_2$.

As in Table~\ref{table:water_vertical}, while the monotonic increase in $P_{acc}$, $E_f$ is non-monotonic with $v_f$; $E_f$ reaches its maximum at $v_f=\sim1.9$ \AA/ps ($N_f=400$) at both solution molalities, yet $E_f$ at 0.44 $m$ is about three-fold higher than at 4 $m$.
The optimal $v_f$, in principle, depends on the solution molality, as it should depend on the timescales of molecular processes in the solutions, such as solvation dynamics, structural relaxation, and ion transport. Similarly, the concentration-dependent $E_f$ is reasonable, since during the exchange the reorganization of non-flying electrolytes is slower at higher salt concentrations due to the larger viscosity of the solution. We found that $P_{acc}$ at the optimal $v_f$ is low at both molalities in comparison to water exchange in liquid water \cite{belloni2019non} reported to exhibit $P_{acc}\approx0.2$ at the optimal $v_f$ of 0.74 \AA/ps.
This implies that the presence of ions at finite salt concentrations further complicates the exchange process using H4D, limiting its efficiency.
We note that the molality of 0.44 $m$ is already not low beyond the valid regime of the Debye-Huckel theory \cite{mcquarrie2000statistical}.

\subsubsection{NaCl-pair exchange}
Ion-pair exchange needs additional considerations (Section~\ref{subsection:trick}), as well as about an order of magnitude longer NEMD than water exchange, including a carefully designed NE potential energy surface and bias techniques.
This section discusses the effects of screening Coulomb interactions, bias functions, and vertical velocity on the efficiency of NaCl-pair exchange in aqueous electrolytes.

\textbf{Effect of screening 4D Coulomb interaction and biasing functions.}\label{subsection:bias_nacl_pair}
\begin {figure*}[htbp]
\includegraphics [width=4.5in] {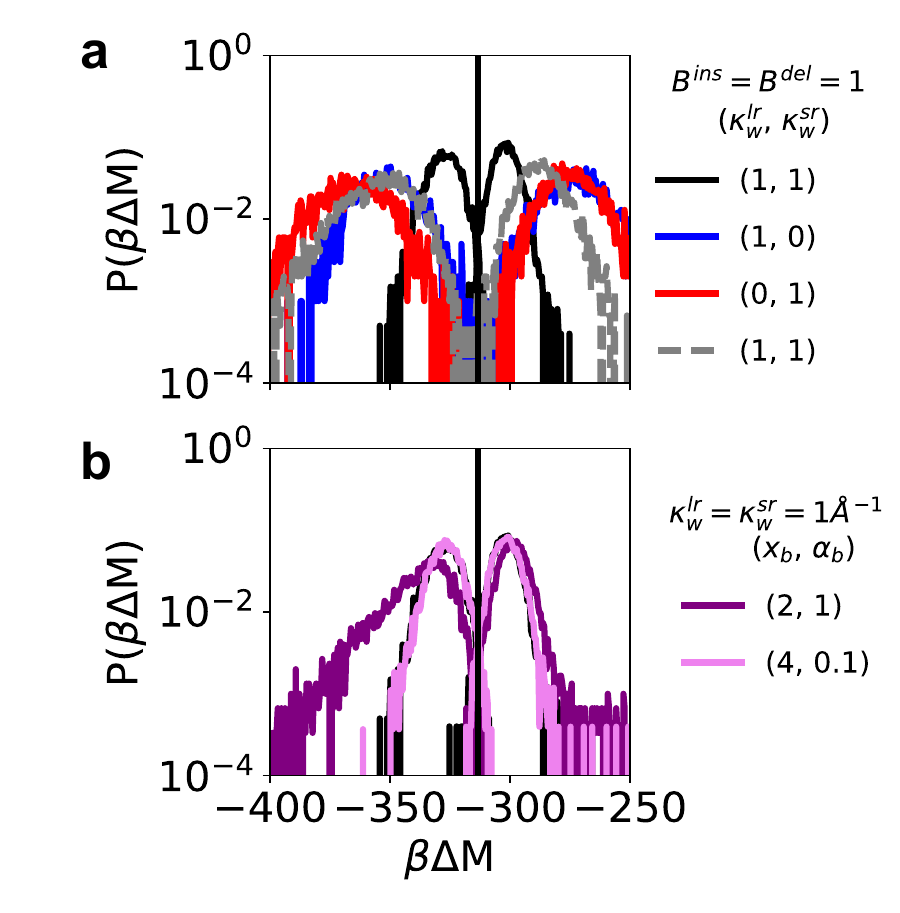}
\caption{
Nonequilibrium work distributions for NaCl-pair exchange (a) with different 4D Coulomb interaction potentials, or (b) with different bimodal biasing functions in aqueous NaCl electrolytes at 1 $m$ concentration.
The vertical black lines in both panels indicate $\beta\mu_{NaCl}=313.4$, calculated using the H4D method.
$\kappa^{lr}_w$ and $\kappa^{sr}_w$ are screening parameters for 4D Coulomb interactions (Eqs~\ref{eq:scr_4d_coul_lr} and~\ref{eq:scr_4d_coul_sr}).
In panel b, a bimodal biasing function (Eq.~\ref{eq:bimodal}) was used for the salt-pair exchange with variable $\alpha^*_b$ and $x^*_b$, while no bias was applied in panel a.
Other parameters for the trial insertion and deletion moves are the same: $\delta t=4$ fs, $w_{max}=3$ \AA$ $, $N_{eq}=2,000$, and $N_f=10,000$ ($v_f=0.075$\AA/ps).
The cut-off distance is 14 \AA$ $ for both LJ and Coulomb interactions during NEMD, except for the grey dotted lines on the top panel (9 \AA$ $ for both interactions).
The calculated $P_{acc}$ are given in Tables~\ref{table:nacl_1m_bias_a} and~\ref{table:nacl_1m_bias_b}.}
\label{nacl_1m_bias}
\end{figure*}
Figure~\ref{nacl_1m_bias} reveals the significant effects of screening 4D Coulomb interactions and the biased distance between flying ions on the nonequilibrium work distributions.
All the distributions for the NaCl-pair exchanges were calculated at fixed $v_f=0.075$\AA/ps.
Figure~\ref{nacl_1m_bias}a clearly shows that without any bias ($B^{ins}=B^{del}=1$) screening both short- and long-range contributions of 4D coulomb interactions significantly increases $P_{acc}$ (Table~\ref{nacl_1m_bias}), resulting in the work distributions that more overlap between trial insertions and deletions; $P_{acc}$ increases about five orders of magnitude at most.
Without such a screening in 4D electrostatic interactions, the NaCl-pair exchange is barely accepted with negligible $P_{acc}$. In principle, we expect $P_{acc}$ to be identical for insertion and deletion with equal numbers of trial moves. However, in practice it is estimated using Eq.~\ref{pacc_theory} with the pre-determined $\beta\mu_{\text{salt}}$ without accepting any trial moves (\textit{i.e.,} at a fixed composition), which may result in numerical errors leading to different values of $P_{acc}$ for insertion and deletion.
Furthermore, we find that increasing the cut-off distance of 4D Coulomb interactions from 9 to 14 \AA $ $ improves the sampling.
A longer cut-off (14 \AA) results in about 2-3 orders magnitude higher $P_{acc}$ than a shorter one (9 \AA). This underlies the importance of the long-range nature of Coulomb interactions in ionic solutions.
The optimal choice of the screening parameters ($\kappa^{lr}_w$ and $\kappa^{sr}_w$) should be system-specific and depend on the salinity.
We find that for the short-range part of 4D Coulomb interactions, screening only the interactions between flying and non-flying particles is more efficient than screening all interactions (not shown here).
\begin{table}
\centering
\begin{tabular}{ |c|c||c| } 
 \hline
 Trial move  & ($\kappa^{lr}_{w}$, $\kappa^{sr}_{w}$ , $r_{cut}$) & $10^2P_{acc}$ \\
 \hline
 \hline
 insertion & (1, 1, 14) & 1.2 \\
 deletion & (1, 1, 14) & 1.8 \\
 \hline
   insertion & (1, 0, 14) & $4\cdot10^{-2}$  \\
 deletion & (1, 0, 14) & $6.9\cdot10^{-3}$ \\
 \hline
  insertion & (0, 1, 14) &  $2.3\cdot10^{-5}$\\
 deletion & (0, 1, 14) &  $3.1\cdot10^{-5}$\\
 \hline
  insertion & (1, 1, 9) & $1.2\cdot10^{-2}$ \\
 deletion & (1, 1, 9) &  $1.5\cdot10^{-1}$\\
 \hline
\end{tabular}
\caption{Estimated acceptance rate $P_{acc}$ (Equation~\ref{pacc_theory}) for NaCl-pair exchange in aqueous NaCl electrolytes at 1 $m$ concentration using the work distributions in Figure~\ref{nacl_1m_bias}a.
$r_{cut}$ is a cut-off distance for Coulomb interactions.
No bias was applied ($B^{ins}=B^{del}=1$).
$\beta\Delta M\in[-500,500]$ and $\beta\mu_{\text{salt}}$ = -313.4 were used in calculating $P_{acc}$.
}\label{table:nacl_1m_bias_a}
\end{table}
\begin{table}
\centering
\begin{tabular}{ |c|c||c| } 
 \hline
 Trial move  & ($x_b$, $\alpha_b$) & $10^2P_{acc}$ \\
 \hline
 \hline
 insertion & (2, 1)  & 0.37 \\
 deletion & (2, 1) &  0.51 \\
 \hline
  insertion & (4, 0.1)  & 1.5 \\
 deletion & (4, 0.1) & 1.3 \\
 \hline
\end{tabular}
\caption{Estimated acceptance rate $P_{acc}$ (Equation~\ref{pacc_theory}) for NaCl-pair exchange in aqueous NaCl electrolytes at 1 $m$ concentration using the work distributions in Figure~\ref{nacl_1m_bias}b.
$\kappa^{lr}_w=\kappa^{sr}_w=-1$ \AA$^{-1}$.
$\beta\Delta M\in[-500,500]$ and $\beta\mu_{\text{salt}}$ = -313.4 were used in calculating $P_{acc}$.}\label{table:nacl_1m_bias_b}
\end{table}

Figure~\ref{nacl_1m_bias}b shows that no significant additional improvement is obtained by introducing a bimodal biasing function (Eq.~\ref{eq:bimodal}) in addition to a proper screening of 4D Coulomb interactions.
Two parameters, namely the sharpness ($\alpha_b$) and mean ion separation ($x_b$), enters in the bias to control the separation distance between flying Na$^+$ and Cl$^-$ ions.
Two particular cases were investigated: one biasing function is to sample only NaCl-pairs quite close to each other ($x_b=2$ \AA, and $\alpha_b=1 $ \AA$^{-2}$), and the other is in various interionic distances ($x_b=4$ \AA, and $\alpha_b=0.1 $ \AA$^{-2}$).
In both cases, both short- and long-range contributions of 4D Coulomb interactions were screened ($\kappa^{lr}_w=\kappa^{sr}_w=1 $ \AA$^{-1}$).
The work distributions with the narrow biasing function ($x_b=2$ \AA, and $\alpha_b=1 $ \AA$^{-2}$) are highly asymmetric: negatively skewed for trial NaCl-pair deletion and positively skewed for trial NaCl-pair insertion.
On one hand, the positive tail for the trial insertions comes from ion pairs with significant overlap. 
In such a case, the early rejection scheme (Section~\ref{subsection:early_rejection}) can help not to waste simulation time by eliminating such highly unfavorable initial configurations before NEMD.
On the other hand, the negative tail for the trial deletions comes from the factor $B^{del}$ (Equation~\ref{metro_del_bias}), significantly deviating from unity; a flying NaCl-pair should be selected from (quasi-)equilibrium configurations in which almost all the ions are away from each other more than ~4 \AA.
Even though the narrow biasing function ($x_b=2$ \AA, and $\alpha_b=1 $ \AA$^{-2}$) deteriorates the sampling with a proper screening of both short- and long-range 4D Coulomb interactions, we found that the narrow biasing function is beneficial in the case of only screening short-range 4D Coulomb interactions (not shown here).
Further discussion regarding the effect of biasing functions is given in Section~\ref{sec:SI_bias_detail} of the SI.

\begin{figure}[htbp]
    \includegraphics [width=3in] {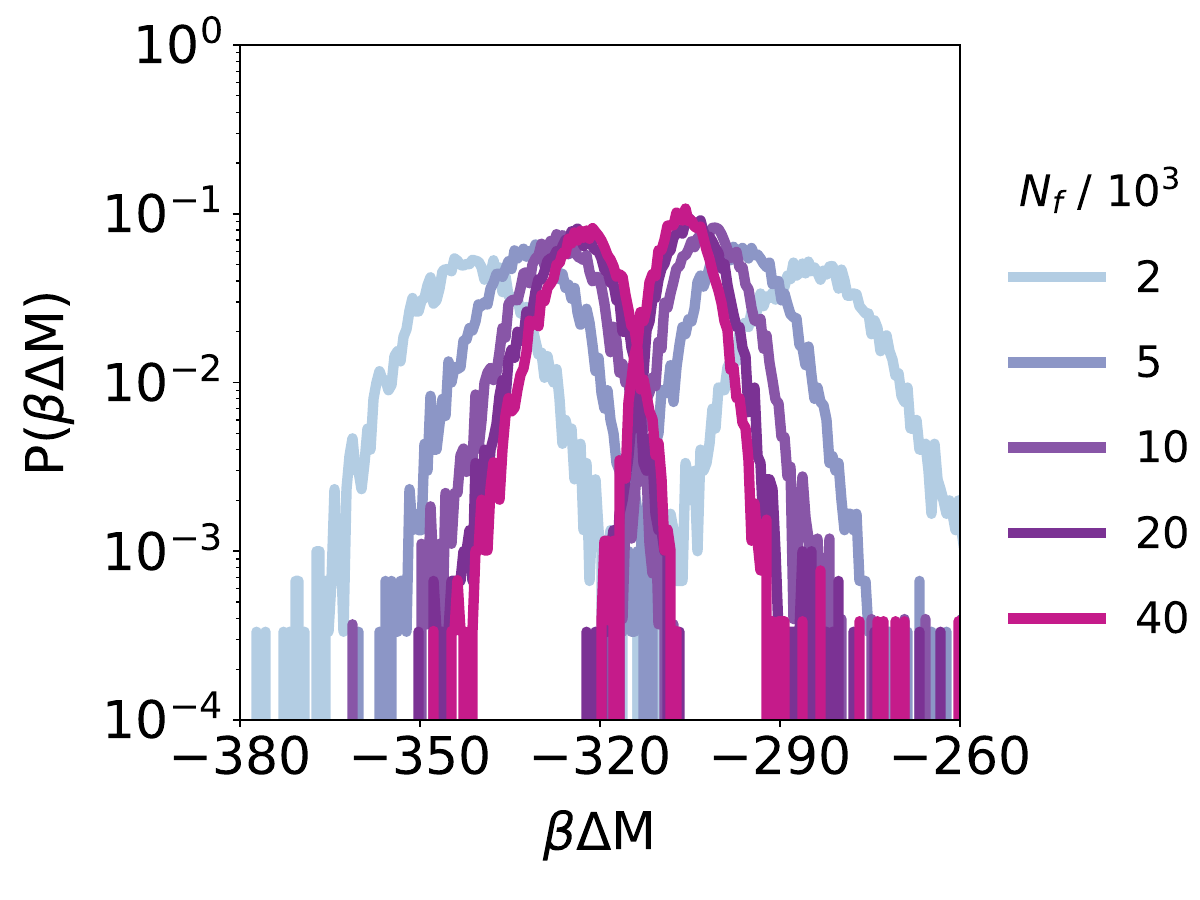}
\caption{Nonequilibrium work distributions for NaCl-pair exchange at several vertical velocities in aqueous NaCl electrolytes at 1 $m$ concentration, obtained during.
No excluded volume was considered, and other parameters for trial insertion and deletion moves are the same: $\delta t=4$ fs, $w_{max}=3$ \AA$ $, $\alpha_b=0.1$ \AA$^{-2}$, $x_b=4$ \AA, and $\kappa^{lr}_w=\kappa^{sr}_w=1 $ \AA$^{-1}$.}\label{nacl_1m_vertical}
\end{figure}

\begin{table}
\centering
    \begin{tabular}{ |c|c|c|c|c| } 
     \hline
     $10^{-3}N_f$  & $10^2P_{acc}$ & $10^2P^{GCMD}_{acc}$ & $10^6E_f$ \\ 
     \hline
     \hline
     2 & $0.013$ & $0.02$ & 0.1 \\ %& 0.00017 \\ 
     5 &  0.38 & 0.26 ($\pm0.06$) & 0.52 \\ %& 0.0042 \\
     10 & 1.5 & 1.2 ($\pm0.1$) & 1.3 \\ %& 0.014 \\
     20 & 2.7 & 2.8 ($\pm0.1$) & 1.4 \\ %& 0.032 \\
     40 & 4.8 & 4.7  ($\pm0.3$) & 1.18 \\ %& 0.059 \\
     \hline
    \end{tabular}
\caption{Acceptance probability ($P_{acc}$) and efficiency ($E_f$) of NaCl-pair exchange at different vertical velocities in aqueous NaCl electrolytes at 1 $m$ concentration.
$P_{acc}$ in this table was calculated using the distributions for trial NaCl-pair insertions in Figure~\ref{nacl_1m_vertical}, while $P^{GCMD}_{acc}$ was obtained from GCMD in the $N_{\text{water}}\mu_{\text{NaCl}}pT$ ensemble.
$\beta\mu_{\text{NaCl}}=-313.4$ and $\beta\Delta M\in[-500,500]$ were used in calculating $P_{acc}$.
The efficiency parameter $E_f$ was calculated using $P^{GCMD}_{acc}$: $E_f=P^{GCMD}_{acc}/N_f$.
Other parameters are given in the caption of Figure~\ref{nacl_1m_vertical}.}\label{tab:nacl_1m_vertical}
\end{table}

\textbf{Effect of the vertical velocity.}
Figure~\ref{nacl_1m_vertical} shows the effect of the vertical velocity on the nonequilibrium work distributions for NaCl-pair exchange at various vertical velocities ($v_f$=0.01875 - 0.375 \AA/ps) at 1 $m$ NaCl concentration.
All the distributions were obtained with the same bimodal biasing function ($\alpha_b=0.1$ \AA$^{-2}$ and $x_b=4$ \AA) and the same screened 4D electrostatic interactions ($\kappa^{lr}_w=\kappa^{sr}_w=1 $ \AA$^{-1}$).
It is evident that NaCl-pair exchange is computationally more demanding than water exchange, as it requires about an order of magnitude longer NEMD for each trial exchange move.
Although all the distributions are nearly Gaussian except for the long tail for trial insertions, they are much broader than the ones for the water exchange (Figure~\ref{water_vertical}), decreasing $P_{acc}$.
As expected, the decreasing $v_f$ monotonically increases $P_{acc}$ (Table~\ref{tab:nacl_1m_vertical}) up to ~0.05 in the range of $v_f$ studied, with narrower distributions and a mean closer to $\mu_{\text{NaCl}}$.
However, $E_f$ reaches its maximum around $v_f=0.0375$ \AA/ps ($N_f=20,000$).
With all the techniques discussed above (Method section~\ref{subsection:trick}), we achieve an acceptance rate of $\sim$3 $\%$ for NaCl-pair exchange in 1 $m$ aqueous NaCl electrolytes at the maximum efficiency.

\begin {figure*}[htbp]
\includegraphics [width=5in] {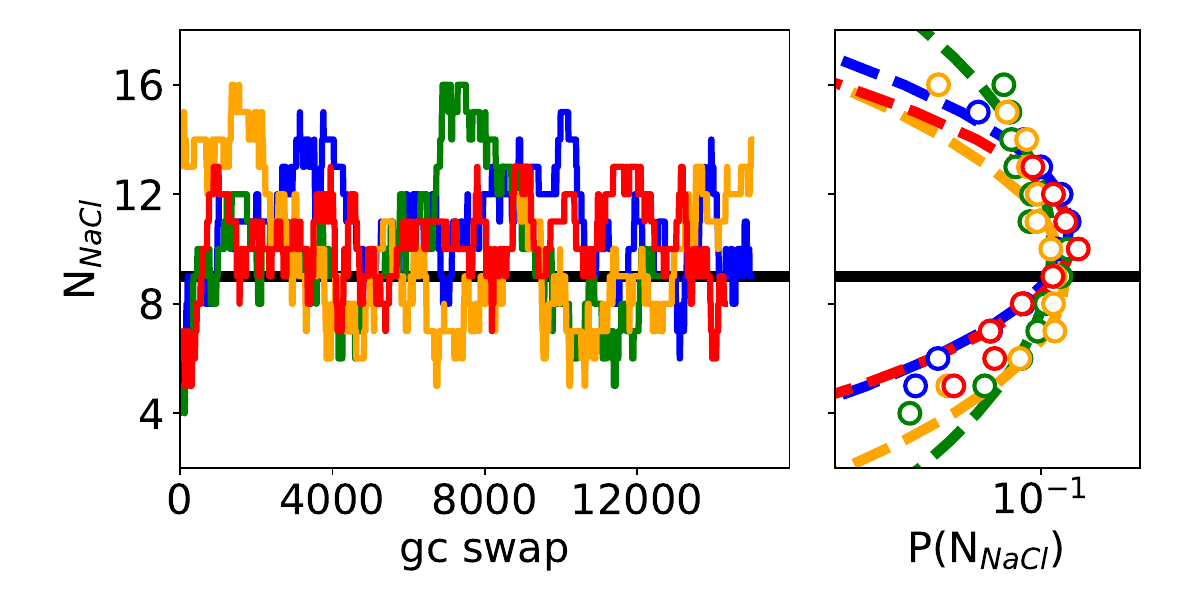}
\caption{Number of NaCl pairs (left) and its distributions (right) during GCMD simulations in the $N_{\text{water}}\mu_{\text{NaCl}}pT$ ensemble.
Different colors represent different independent MD runs.
The black horizontal lines in both panels indicate the number ($N_{\text{NaCl}}=9$) of NaCl pairs in calculating $\beta\mu_{\text{NaCl}}=313.4\pm0.3$ in the $N_{\text{water}}N_{\text{NaCl}}pT$ ensemble.
During NEMD, the cut-off distance is 14 \AA$ $ for both LJ and Coulomb interactions, $\delta t=4$ fs, $w_{max}=3$ \AA$ $, $N_{eq}=2,000$, and $N_f=10,000$.
A bimodal biasing function was used with $\alpha_b=0.1$ \AA$^{-2}$, $x_b=4$ \AA, and $\kappa^{lr}_w=\kappa^{sr}_w=1 $ \AA$^{-1}$.
A small excluded volume ($V_{ex}=0.125$\AA$^3$) was also applied for early rejection (See Section~\ref{subsection:early_rejection}).
Dotted lines on the right panel represent Gaussian distributions from each MD run with the measured average and variance.}
\label{nacl_1m_gcmc_salt}
\end{figure*}

\textbf{Grand-canonical MD simulations.}
Figure~\ref{nacl_1m_gcmc_salt} shows the fluctuations of the number of NaCl pairs during GCMD simulations  with H4D at $v_f=0.075$ \AA/ps in the $N_{\text{water}}\mu_{\text{NaCl}}pT$ ensemble.
In all cases, $N_{\text{water}}$ was fixed to 500, and $\beta\mu_{NaCl}$ was fixed to 313.4, which was computed in the $N_{\text{water}}N_{\text{NaCl}}pT$ ensemble using the H4D method with $N_{\text{water}}=500$ and $N_{\text{NaCl}}=9$, and little different from the value (314.5) reported in Ref~\onlinecite{mester2015mean} using the thermodynamic integration method.
In all four independent runs, $N_{\text{NaCl}}$ oscillates around and converges to the similar value used in calculating $\beta\mu_{NaCl}$, exhibiting $\langle N_{\text{NaCl}}\rangle_{GC}=10.0\pm0.4$, and $\sqrt{\langle(\delta N_{\text{NaCl}})^2)\rangle_{GC}}=2.0\pm0.3$.
We note that the statistical error in the computed $\beta\mu_{NaCl}$, with no finite-size correction, in the $N_{\text{water}}N_{\text{NaCl}}pT$ ensemble may lead to the small deviation of $\langle N_{\text{NaCl}}\rangle_{GC}$ from  $N_{\text{NaCl}}=9$; 
we found that the statistical error in $N_{\text{NaCl}}$, estimated by back-propagation of the statistical error (0.3) in $\beta\mu_{NaCl}$, is 1.2 with the computed $\langle(\delta N_{\text{NaCl}})^2)\rangle_{GC}$.
The GCMD simulations achieve the acceptance rate of $\sim$$1 \%$ at $v_f=0.075$ \AA/ps.

The Kirkwood-Buff theory \cite{kirkwood1951statistical,kusalik1987thermodynamic} allows for calculating the osmotic compressibility (Equation~\ref{osmotic}) from salt-density fluctuations using ion-ion structure factors.
In this procedure, no finite-size corrections are needed for the ion-ion structure factors computed in the $N_{\text{water}}\mu_{\text{NaCl}}pT$ ensemble, while they are essential in the $N_{\text{water}}N_{\text{NaCl}}pT$ ensemble \cite{belloni2019non}.
We found all three ion-ion structure factors converge to the same value ($\chi_{osmotic}=0.48\pm0.05$) close to the value 0.5 expected from the Debye-Hückel theory \cite{kusalik1987thermodynamic} that reduces to the ideal gas prediction for sufficiently dilute eletrolytes (See the SI). Furthermore, Equation~\ref{osmotic} also allows us to estimate the water-density fluctuation $\sqrt{\langle( \rho_{\text{water}}-\langle\rho_{\text{water}}\rangle)^2\rangle}$ in the $\mu_{\text{water}}N_{\text{NaCl}}pT$ ensemble, which turns out to be huge, being about 200 with $\chi_{osmotic}=0.5$.
The large fluctuation implies that a long GCMD simulation is needed to be performed for a correct sampling in the $\mu_{\text{water}}N_{\text{NaCl}}pT$ ensemble.

\section{Conclusions}
Despite their significance in many applications, GCMD simulations are still computationally demanding so that their use remains limited in practice.
In this work, we implemented in LAMMPS a promising hybrid NEMD/MC method, called H4D, which utilizes a vertical dimension to facilitate particle exchange by alleviating initial steric and electrostatic clashes.
The H4D method is conceptually simple and requires minimal code changes for a conventional MD simulation.
With our implementation, we showed that GCMD simulations with H4D efficiently describes a system in an open environment in a condensed phase, such as ionic solutions.

The H4D is a finite switching method, interpolating between instantaneous and infinitely slow exchange; the acceptance rate increases with slower exchange that needs a longer NEMD simulation.
Thus, one can optimize the process, which should be system-specific, using the various ingredients as discussed in this work, including altitude, vertical velocity, screening 4D electrostatic interactions, and bias, through analyzing the nonequilibrium work distributions.
Our investigation underlies the crucial role of long-range electrostatic interactions, and their proper screening can significantly enhance the efficiency of ion-pair exchange in electrolyte solutions; the H4D enhances the efficiency of salt-pair exchange about four orders of magnitude, compared to the conventional MC.
Further,  at its maximum efficiency, the H4D achieves the acceptance rate of $\sim3\%$ for NaCl-pair exchange in aqueous solutions at 1 $m$ concentration, which is manageable with massively paralleled computation in LAMMPS.

We also investigated the effect of biased distances between flying ions.
It showed no further benefit in the efficiency with a proper screening of electrostatic interactions; in a case of the bias with a flying-ion pair too close to each other, the nonequilibrium work distributions are skewed in unfavorable ways.
For further enhanced efficiency, a better bias needs to be designed to sample the skewed distributions in a favorable way.

GCMD simulations using H4D is generic and our implementation in LAMMPS is flexible enough for applications to other bulk systems in an open environment.
Furthermore, the extension to a confined systems is straightforward with a few considerations such as the initial positions of flying particles along the non-periodic dimension.
In such a case, a similar bias technique can be applied to avoid steric clashes with a implicit or explicit wall, generating the initial positions at the center of the confined system.

\section*{Acknowledgement}
This project received funding from the European Research Council under the European Union’s Horizon 2020 research and innovation program (grant agreement no. 863473).

\section*{Data Availability Statement}\label{sec:DataAvail}
Our implementation of H4D is freely available at \href{https://github.com/Jeongmin0658/h4d_lammps}{https://github.com/Jeongmin0658/h4d\_lammps}.
All the data presented in this work will be provided upon reasonable requests.

\newpage
\newpage

\newcommand{\thetitle}{Supporting Information for\\Grand-canonical molecular dynamics simulations powered by a hybrid nonequilibrium MD/MC method via the 4$^{th}$ dimension: Implementation in LAMMPS and applications to bulk electrolytes}
%\begin{suppinfo} 
\section*{Appendix}
\label{suppinfo}

\renewcommand{\thesection}{S\arabic{section}}
\setcounter{section}{0}
\renewcommand{\thefigure}{S\arabic{figure}}
\setcounter{figure}{0}
\renewcommand{\thetable}{S\arabic{table}}
\setcounter{table}{0}
\renewcommand{\theequation}{S\arabic{equation}}
\setcounter{equation}{0}

\section{Chemical potential calculation: Crooks theorem and BAR method}\label{subsubsec:cft_bar}
Similarly to the Widom method \cite{widom1963some}, H4D \cite{belloni2019non} can be also used to calculate chemical potential.
H4D is a finite-time switching method that interpolates between the Widom method, instantaneously creating or deleting a particle, and thermodynamic integration that follows a quasi-equilibrium path \cite{dellago2013computing}.
In calculating chemical potentials, the difference in total energy is sampled, yet no trial move should be accepted, keeping the constant number of the species of interest.

Chemical potentials (and free energies, in general) can be computed using the Crooks fluctuation theorem (CFT) \cite{crooks1999entropy}, which connects the work distributions of trial insertion and deletion moves:
\begin{equation}\label{cft}
\frac{P_{ins}(\Delta M)}{P_{del}(\Delta M)}=\beta(\Delta M-\mu).
\end{equation}
We note that the sign of the nonequilibrium work in both trial moves follows the direction of trial insertions as in Ref~\onlinecite{belloni2019non}.
One well-known application of the CFT is the measurement fo the folding free energy of RNA hairpin via pulling experiment \cite{collin2005verification}. 

According to Equation~\ref{cft}, $P_{ins}$ and $P_{des}$ should intersect each other at $\Delta M=\mu$.
Thus, in order to estimate $\mu$ using the CFT, one needs to find the intersection point. 
In this work, we calculated the chemical potential using the BAR method \cite{bennett1976efficient,shirts2003equilibrium}:
\begin{widetext}
\begin{equation}\label{bar_eq}
    \sum_{i=1}^{N_{ins}}\frac{1}{1+\exp[{\beta(C+\Delta M - \mu)}]}
    =\sum_{j=1}^{N_{del}}\frac{1}{1+\exp[{-\beta(C+\Delta M - \mu)}]},
\end{equation}
\end{widetext}
where $C=\ln(N_{ins}/N_{del})/\beta$, and $N_{ins}$ and $N_{del}$ are the number of configurations sampled from trial insertion and deletion, respectively.
The equation above was derived using the maximum likelihood arguments for the BAR method \cite{shirts2003equilibrium}, and numerically solved using Newton's method.
We note that another way of finding the intersection is by fitting the work distributions using polynomial or Gaussian functions as was suggested in Ref~\onlinecite{belloni2019non}. 
One should keep in mind that the chemical potential depends on the choice of $\Lambda_s$; a different choice of $\Lambda_s$ shifts the distribution functions horizontally, resulting in a different intersection point.

\section{Gibbs-Duhem equation: Thermodynamic consistency test of calculated chemical potential}\label{sec:SI_gde}
To validate our calculated chemical potential using H4D, Gibbs-Duhem equation (GDE) was tested with independent calculations of solvent and salt chemical potential in a wide range of solution molality for the LJ model electrolytes.

For a binary mixture (such as aqueous NaCl solutions), GDE with activity $a_i$ and mole fraction $x_i$ for $i$-species ($i \in \{s,w\}$) becomes:
\begin{equation}
\begin{split}
x_sd\ln(a_s)+x_wd\ln(a_w)=0.
\end{split}
\label{gde_binary}
\end{equation}

For an electrolyte system, the salt chemical potential ($\mu_s$) is expressed as a function of $m$ as follows \cite{mouvcka2013molecular}:
%\begin{widetext}
\begin{equation}
\begin{split}
& \beta\mu_s(m) =\beta\mu^\dagger_s + 2 \ln m \\
& + 2\ln10\cdot\bigg(
-\frac{A\sqrt{m}}{1+B\sqrt{m}}+Em+Cm^2+Dm^3
\bigg),
\end{split}\label{chem_salt}
\end{equation}
%\end{widetext}
where $D$, $C$, and $E$ are fitting parameters to take into account the deviation from the DH limiting law. 
$\mu^\dagger_s$ is also a fitting parameter, which is salt chemical potential at infinite dilution. 
The DH parameter are predetermined: $A=\frac{1}{\ln10}l^*_B\sqrt{2\pi l^*_B\cdot \frac{\rho^*}{1000}}$, and $B=a^*\sqrt{8\pi l^*_B\cdot \frac{\rho^*}{1000}}$ with $\rho^*$ being a pure water density.
Then, from the GDE, the water chemical potential ($\mu_w$) is expressed as follows:
%\begin{widetext}
\begin{equation}
\begin{split}
\beta\mu_w(m) & =\beta\mu^0_w- 2m\frac{M_w}{1000}\\
&-\ln 10\cdot \frac{M_w}{1000}\bigg(
Em^2+\frac{4C}{3}m^3+\frac{3D}{2}m^4 \\
& +\frac{2A}{B^3+B^4\sqrt{m}}
+\frac{4A\ln(B\sqrt{m}+1)}{B^3} \\
& -\frac{2A\sqrt{m}}{B^2}-\frac{2A}{B^3}
\bigg).
\end{split}\label{chem_solv}
\end{equation}
%\end{widetext}
As discussed above, the only free parameter for $\mu_w(m)$ is $\mu^0_w$, water chemical potential of pure water, since the parameters ($D$, $C$, and $E$) can be determined from $\mu_s(m)$.

\begin {figure*}[htbp]
\includegraphics [width=3in] {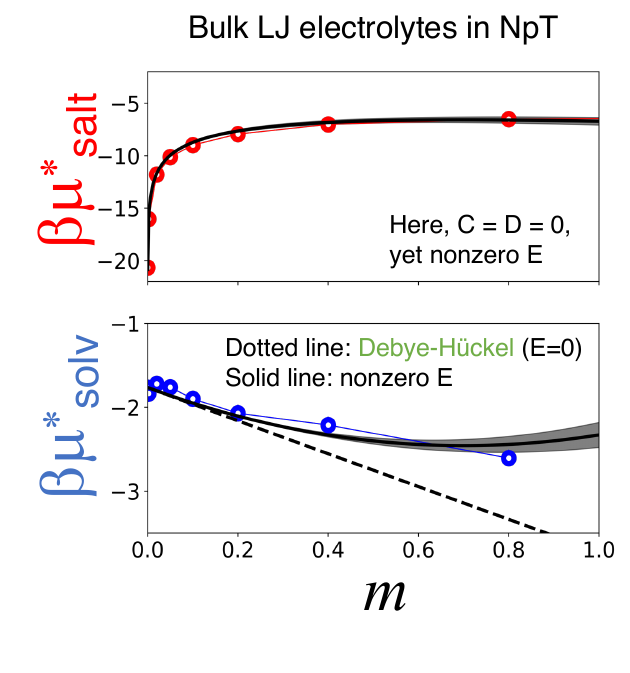}
\caption{Testing Gibbs-Duhem relation for the model LJ electrolytes with calculated chemical potential of solvent ($\mu_{\text{solv}}$) and mean chemical potential of an ion pair ($\mu_{\text{salt}}$) in a range of solution molality ($m$).
Solid and dotted lines are drawn using Equation~\ref{chem_salt} and ~\ref{chem_solv}.
Grey shadow region on the bottom panel represents the 95$\%$ confidence interval.}\label{lj_gde}
\end{figure*}

Figure~\ref{lj_gde} shows the chemical potential of both species in a range of solution molality ($m$), calculated in the $N_{\text{solv}}N_{\text{salt}}pT$ ensemble.
In this range of $m$, $\mu_{\text{salt}}(m)$ is well described with the extended Debye-Hückel and the linear contributions (Equation~\ref{chem_salt}) with no higher-order terms considered.
With the pre-determined DH parameters, the only parameter to fit $\mu_{\text{salt}}(m)$ (solid black line on the top panel) is the coefficient of the linear term, which is needed to describe the plateau region ($0.2\lesssim m \lesssim0.8$) after an initial steep increase at low $m$.

The bottom panel of Figure~\ref{lj_gde} clearly shows the calculated chemical potential satisfies the GDE (Equation~\ref{gde_binary}).
The solid black line, consistent with the calculated $\mu_{\text{solv}}(m)$, is not a fit result, but the result of the GDE with the fit parameter for $\mu_{\text{salt}}(m)$, and the chemical potential of the pure LJ solvent at $m=0$.
Thus, from this thermodynamic consistency, we can validate the calculated chemical potentials via the H4D.

\section{Simple, early rejection scheme for ion-pair exchange to save computational time via instantaneous GCMC}\label{subsection:early_rejection}
In the case of ion-pair exchange, the acceptance rate significantly depends on the interionic distance between the two flying ions, with two competing factors.
On the one hand, a flying-ion pair with too large a separation likely leads to rejection since the pair may perturb the system too strongly.
On the other hand, a flying-ion pair with a significant steric repulsion with each other also likely leads to rejection.
Further, such a flying-ion pair with a large overlap could raise a numerical instability in the following equilibrium MD, in addition to its rare acceptance.
This section discusses a scheme to avoid wasting a computational time to run a NEMD in H4D for such a configuration that contains a flying-ion pair with a large overlap.

This approach introduces a small enough excluded volume, $V_{ex}$, in which overlaps between the two flying ions are non-physical.
In this work, we define the excluded volume as a cubic region, centered around the flying cation.
Even with a proper bias function, such a flying-ion pair can be prepared in a trial insertion move, while it is rarely prepared in a trial deletion move.
For both trial insertion and deletion moves, an essential part is to satisfy the detailed balance, taking into account the excluded volume.
To do so, instead of discarding such a configuration in which flying ions touch each other in $V_{ex}$, a conventional MC that instantaneously creates or destructs an ion-pair is employed, following the idea of cavity-biased MC \cite{mezei1980cavity}.

The pre-assigned size of $V_{ex}$ centered around the flying cation determines the probability, $p_{NE}$, of employing NEMD: $p_{NE}=1-V_{ex}/V$, where $V$ is the volume of a simulation box. 
Accordingly, the probability, $p_{cv}$, of employing an instantaneous MC (\textit{i.e.,} no NEMD, but potential energy evaluations for a trial move) is simply $p_{cv}=1-p_{NE}=V_{ex}/V$.
We note that with a small enough $V_{ex}$, $p_{cv}$ is quite small ($p_{cv} < 0.001$ in most cases in this work).
In order to satisfy the detailed balance, a trial removal move must apply the same probabilities that NEMD is carried out with the probability of $p_{NE}$, or an instantaneous conventional MC move with the probability of $p_{cv}$; 
this can be done by drawing a random number for each trial removal move.
We note that the detailed balance must be satisfied even by simply rejecting such an ion pair if the probability of finding an ion pair within $V_{ex}$ is strictly zero (\textit{e.g.}, hard spheres).
This rejection scheme requires no changes in the Metropolis acceptance rule, except that a conventional MC move should employ a standard Metropolis acceptance rule without the kinetic energy part.

\begin{table*}
\centering
\begin{tabular}{ |c|c|c||c|c|c|c|c| } 
 \hline
 $m$ & Trial move & $N_f$ & $\langle\beta\Delta M\rangle-\beta\mu_{\text{water}}$ & $\langle(\delta\beta\Delta M)^2\rangle$ & $\alpha_2$ & $10^2P_{acc}$ & $10^4E_f$ \\ 
 \hline
 \hline
 0.44 & Insertion & 100 & 28 & 210 & 0.63 & 0.42 & 0.42 \\ 
0.44 & Insertion & 200 & 17 & 66 & 0.38 & 1.0 & 0.52 \\ 
0.44 & Insertion & 400 & 11 & 31 & 0.17 & 2.7 & 0.68 \\ 
0.44 & Insertion & 800 & 8.5 & 20 & 0.18 & 4.6 & 0.57 \\ 
0.44 & Insertion & 1600 & 7.0 & 16 & 0.12 & 6.5 & 0.41 \\ 
 \hline
0.44 & Deletion & 100 & -18 & 39 & 0.048 &  0.28 & 0.28 \\
0.44 & Deletion & 200 & -15 & 37 & 0.079 &  0.90 & 0.45 \\ 
0.44 & Deletion & 400 & -12 & 32 & 0.063 &  2.2 & 0.54 \\ 
0.44 & Deletion & 800 & -10 & 28 & 0.093 &  4.3 & 0.54 \\ 
0.44 & Deletion & 1600 & -8.2 & 26 & 0.13 &  7.1 & 0.44 \\ 
 \hline
  \hline
4 & Insertion & 100 & 35 & 320 &  1.3 & 0.18 & 0.18 \\
4 & Insertion & 200 & 23 & 110 &  0.45 & 0.42 & 0.21\\
4 & Insertion & 400 & 16 & 59 &  0.11 & 0.91 & 0.23 \\
4 & Insertion & 800 & 13 & 38 &  0.68 & 1.6 & 0.19 \\
4 & Insertion & 1600 & 12 & 28 &  -0.007 & 1.9 & 0.12 \\
 \hline
4 & Deletion & 100 & -22 & 57 &  -0.032 & 0.15 & 0.15 \\
4 & Deletion & 200 & -19 & 49 &  0.012 & 0.40 & 0.20 \\
4 & Deletion & 400 & -16 & 43 &  0.009 & 0.87 & 0.22 \\
4 & Deletion & 800 & -14 & 39 &  0.051 & 1.6 & 0.20 \\
4 & Deletion & 1600 & -13 & 37 &  0.091 & 2.1 & 0.13 \\
 \hline
\end{tabular}
\caption{Characterization of the work distributions in Figure~\ref{water_vertical} for water exchange at different vertical velocity at 0.44 $m$ NaCl (top) and 4 $m$ NaCl (bottom) molalities.
Water chemical potentials ($\beta\mu_{\text{water}}=-15.34$ and -15.51 for 0.44 and 0.4 $m$, respectively) were taken from Ref~\onlinecite{mester2015mean}.
The efficiency parameter $E_f=P_{acc}/N_f$, and $\alpha_2$ is the non-Gaussian parameter.}
\label{table:water_vertical}
\end{table*}

\section{Effect of a bias on NaCl-pair exchange in 1 $m$ NaCl (aq)}\label{sec:SI_bias_detail}
\begin {figure*}[htbp]
\includegraphics [width=6.5in] {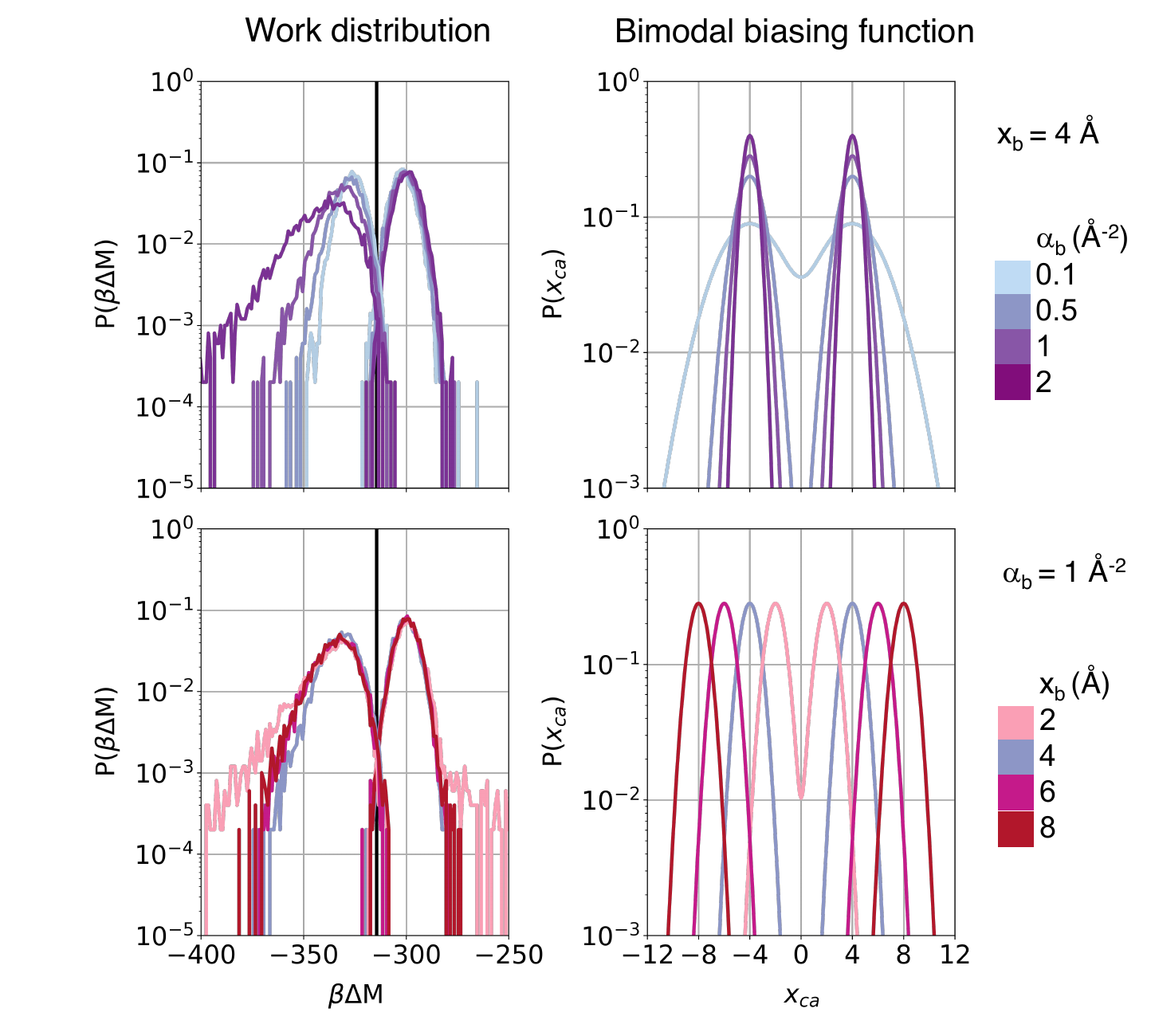}
\caption{Work distributions (left column) of NaCl-pair exchange with different bimodal biasing functions (right column) at the same vertical velocity in aqueous NaCl electrolytes at 1 $m$ concentration.
Other parameters are the same: $\delta t=4$ fs, $w_{max}=3$ \AA$ $, and Nsteps=10,000.}
\label{SI_nacl_1m_bias}
\end{figure*}
\begin {figure*}[htbp]
\includegraphics [width=6.5in] {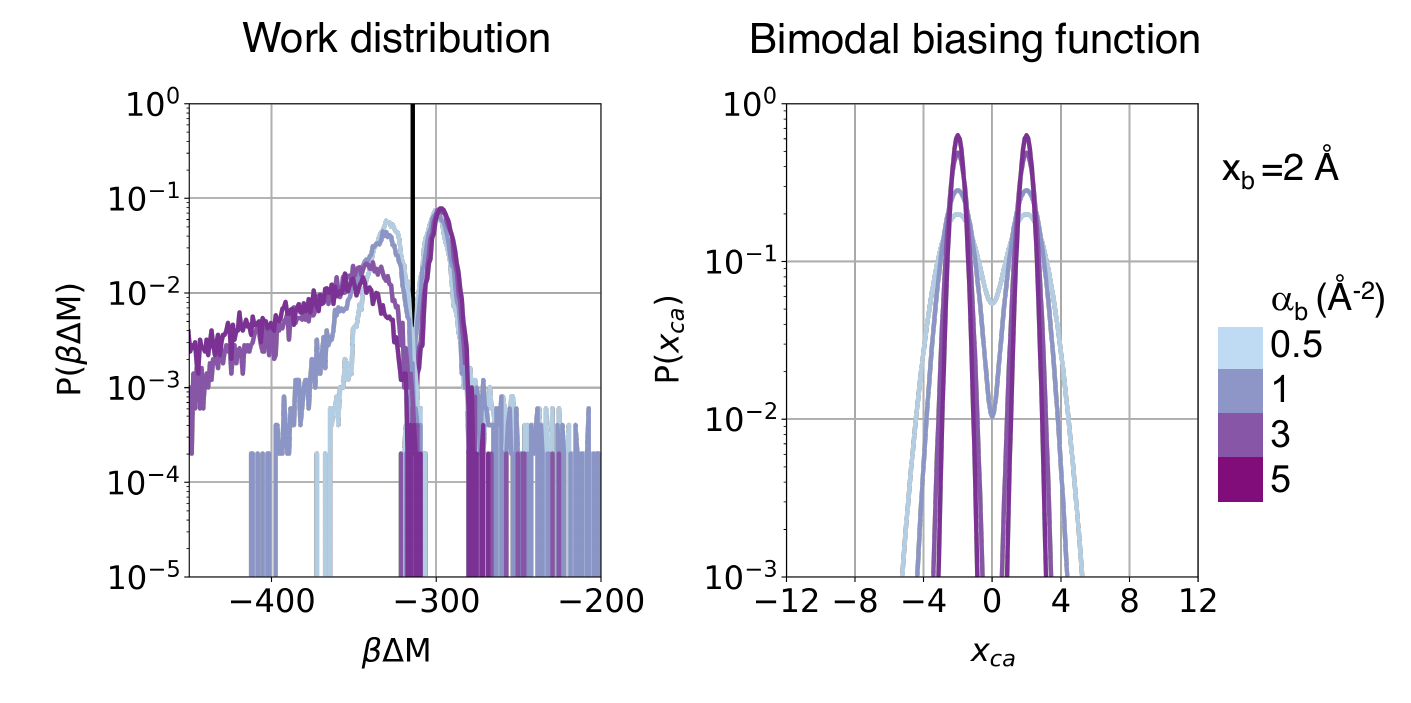}
\caption{Work distributions (left column) of NaCl-pair exchange with different bimodal biasing functions (right column) at the same vertical velocity in aqueous NaCl electrolytes at 1 $m$ concentration.
$x_b$ is fixed to be 2 \AA, yet $\alpha_b$ varies between 0.1 and 5 \AA$^{-2}$.
Other parameters are the same: $\delta t=4$ fs, $w_{max}=3$ \AA$ $, and Nsteps=10,000.}
\label{SI_nacl_1m_bias_xb2}
\end{figure*}

Figure~\ref{SI_nacl_1m_bias} and Table~\ref{table:SI_1m_salt_bias} show the work distributions for NaCl-pair exchange with a bimodal biasing function with two parameters at play, including the sharpness ($\alpha_b$) and mean ion separation ($x_b$), to control the separation distance between flying Na$^+$ and Cl$^-$ ions.
All the distributions are calculated at fixed $v_f$.
It is immediately apparent that a biasing function is critical in determining the shape of the work distribution and, thereby, the efficiency of ion-pair exchange.
The top row shows the effect of $\alpha_b$ on the work distribution: with $x_b=4$ \AA, the efficiency of NaCl-pair exchange is expected to decrease with increasing $\alpha_b$ based on $R_{ov}$.
Unlike the work distributions for water exchange, the distributions for NaCl-pair exchange are highly non-Gaussian with tails.
On the one hand, a too narrow biasing function with a larger $\alpha_b$ (dark purple) leads to an asymmetric work distribution with a negative skewness for trial NaCl-pair deletion moves.
On the other hand, a smaller $\alpha_b$ (sky blue) leads to a relatively symmetric work distribution with positive skewness for NaCl-pair trial insertion moves, which is due to large overlap between the flying ions.
The same conclusions are drawn with the work distributions with $x_b=4$ \AA $ $ (Figure~\ref{SI_nacl_1m_bias_xb2} in SI).
We note that the skewed work distribution is desired, yet the other way around: such a distribution is preferred, being skewed around the intersection.

The bottom row in Figure~\ref{SI_nacl_1m_bias}, on the other hand, shows the effect of $x_b$ at a fixed $\alpha_b=1$ \AA$^{-2}$: $R_{ov}$ is non-monotonic with $x_b$.
First, with increasing $x_b$, the skewness significantly decreases, indicating that the long tails in the work distribution at $x_b=2$ \AA $ $ (Figure~\ref{SI_nacl_1m_bias_xb2} in SI) come from the configurations with a NaCl flying-ion pair that the ions are too close to each other.
However, the decreased skewness does not always lead to the increased $R_{ov}$; some close contact ion pairs could result in a favorable configuration to be accepted.
We note that no early rejection scheme is applied in work distributions calculation is applied.
The early rejection scheme could work to alleviate the long tails with excluded volume to eliminate flying-ion pairs with significant overlap.
Thus, a small enough $x_b$ benefits the efficiency of ion-pair exchange with more close contact flying-ion pairs, and at the same time with the early rejection scheme to exclude such flying-ion pairs of high overlap with each other.
We found a bimodal biasing function for aqueous solution at 1 $m$ works well with $\alpha_b=0.1$ \AA$^{-2}$ and $x_b=4$ \AA, exhibiting the highest $R_{ov}$ among the parameters studied here.

\begin{table}[ht!]
\centering
\begin{tabular}{ |c|c|c||c|c|c| } 
 \hline
 Trial move & $x_b$ (\AA) & $\alpha_b$ (\AA$^{-2}$) & $\alpha_2$ & skewness & $R_{ov}$ \\ 
 \hline
 \hline
Insertion & 2 & 0.5 & 139 & 19 & 0.0038 \\ 
Insertion & 2 & 1 & 470 & 35 & 0.0022 \\
Insertion & 2 & 3 & 526 & 31 & 0.0006\\
Insertion & 2 & 5 & 0.12 & 0.2 & 0.0002\\
 \hline
Deletion & 2 & 0.5 & 0.34 & -0.66 & 0.0078 \\ 
Deletion & 2 & 1 & 0.88 & -1.3 & 0.0028 \\
Deletion & 2 & 3 & 1.7 & -1.9 & 0.0014 \\
Deletion & 2 & 5 & 1.0 & -1.6 & 0.001 \\
 \hline
 \hline
Insertion & 4 & 0.1 & 1032 & 51.2 & 0.0074 \\ 
Insertion & 4 & 0.5 & 0.018 & 0.093 & 0.0022 \\ 
Insertion & 4 & 1 & 0.016 & 0.12 & 0.0014 \\
Insertion & 4 & 2 & 0.018 & 0.062 & 0.0012\\
 \hline
Deletion & 4 & 0.1 & 0.057 & -0.19 & 0.014 \\ 
Deletion & 4 & 0.5 & 0.084 & -0.30 & 0.0074 \\ 
Deletion & 4 & 1 & 0.26 & -0.53 & 0.0062 \\
Deletion & 4 & 2 & 0.43 & -0.89 & 0.0034 \\
 \hline
 \hline
Insertion & 2 & 1 & 470 & 35 & 0.0022 \\
Insertion & 4 & 1 & 0.016 & 0.12 & 0.0014 \\
Insertion & 6 & 1 & 0.054 & 0.035 & 0.0028 \\
 \hline
Deletion & 2 & 1 & 0.88 & -1.3 & 0.0028 \\
Deletion & 4 & 1 & 0.26 & -0.53 & 0.0062 \\
Deletion & 6 & 1 & 0.046 & -0.48 & 0.003 \\
\hline
\end{tabular}
\caption{Characterization of the work distributions for NaCl-pair exchange with a bimodal biasing function of various parameters in aqueous NaCl electrolyte at 1 $m$ concentration in Figure~\ref{SI_nacl_1m_bias}.
Vertical velocity is fixed at 0.075 \AA/ps with nsteps=10,000.}
\label{table:SI_1m_salt_bias}
\end{table}

Unlike the work distributions for water exchange, the distributions for NaCl-pair exchange are highly non-Gaussian with long tails.
On the one hand, a too narrow biasing function with a larger $\alpha_b$ (dark purple) leads to an asymmetric work distribution with a negative skewness for trial NaCl-pair deletion moves.
On the other hand, a smaller $\alpha_b$ (sky blue) leads to a relatively symmetric work distribution with positive skewness for NaCl-pair trial insertion moves.

\section{Time-dependent altitude schedules of flying particles in H4D}\label{sec:SI_altitude}
\subsection{Other altitude schedules different from a constant-velocity one}
We introduce two other candidates for an altitude schedule to facilitate the sampling (\textit{i.e.}, achieving higher acceptance ratio in MC at the same computational cost in NEMD).
The proposed altitude schedules ($w_v$ and $w_x$) include a time-dependent altitude velocity.
The altitude protocol again should be predetermined, even though its velocity is variable.
The first one, $w_v(t)$ is as follows: 
\begin{equation} \label{concave}
w_v(t)=
\begin{cases}
w_{\text{max}}(1-x_i^n) \text{ for insertion},\\
w_{\text{max}}(1-x_d)^n \text{ for deletion},
\end{cases}
\end{equation}
where $n\ge1$, $x_i=\frac{t-t_i}{t_f-t_i}$, and $x_d=\frac{t_f-t}{t_f-t_i}$.
Both $x_i$ and $x_d$ lie between 0 and 1.
In this case, for a trial insertion, a flying molecule is inserted slowly at earlier times and rapidly at later times.
On the contrary, for a trial removal, a flying molecule is removed at a faster rate at earlier times than later ones.
The second protocol, $w_x(t)$, does exactly the opposite to the first one, $w_v(t)$, as follows:
\begin{equation} \label{convex}
w_x(t)=
\begin{cases}
w_{\text{max}}(1-x_i)^n \text{ for insertion},\\
w_{\text{max}}(1-x_d^n) \text{ for deletion}.
\end{cases}
\end{equation}
In both proposed protocols, the exponent, $n$, determines their curvature in a $w-t$ plane: a schedule with larger $n$ deviates more from the constant-velocity one. 
When $n=1$, both protocols return to the constant-velocity one in the main text.
For a water exchange move, there is no enhancement in sampling with the proposed protocols, but the constant-velocity schedule seems to be the best among three.
We also note that in the case of pure water, the efficiency of water exchange is enhanced by a factor of two by using a faster vertical velocity below $w'_{max}<w_{max}$ ($w'_{max}=1$ \AA $ $ for instance).

\subsection{Different $w_{max}$ for different ion species.}
Even though trial moves using the H4D work well in most cases, they could not avoid a rare but large overlap between flying ions, if they are supposed to depart from the same altitude in the beginning of a trial insertion move.
A simple remedy could be assigning different $w_{max}$ for a cation and an anion, so there is no serious overlap between any particles in 4D.
In this method, one has to assign a different altitude velocity accordingly, since both ions should simultaneously arrive at 3D space with $w=0$.
Further, in a trial removal move, the flying ions should arrive at different altitude as assigned in a trial insertion move.
For instance, a smaller ion (e.g., Na$^+$ in NaCl) is at a smaller $w_{max}=w_{max,c}$, while the other (e.g., Cl$^-$ in NaCl) is at a larger $w_{max}=w_{max,a}=w_{max,c}+0.5(\sigma_{Na}+\sigma_{Cl})f_w$ with a non-negative $f_w$.
Then, in case of a constant-velocity schedule, the altitude velocity, $v_{w,a}$, of Cl$^-$ is larger than the one, $v_{w,c}$, of Na$^+$: $v_{w,a}$=$v_{w,c}\cdot r_w$ with $r_w=w_{max,a}/w_{max,c}$.
In this approach, there is no change in expression of the Metropolis acceptance criterion.

\section{Connection to the Kirkwood-Buff theory}
\begin {figure}[htbp]
\includegraphics [width=3in] {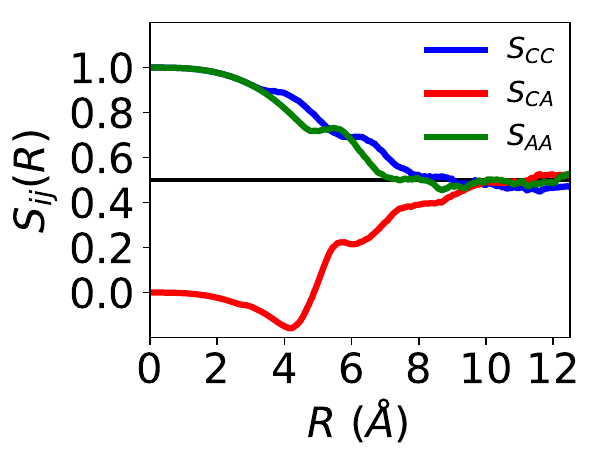}
\caption{"Running" ion-ion structure factors, $S_{ij}(R)$ in aqueous NaCl electrolytes at 1 $m$ concentration: $S_{CC}$ between Na cations, and $S_{AA}$ between Cl anions, and $S_{CA}$ for between Na and Cl ions. 
A black solid line represents $\chi_{osmotic}=0.5$, expected from the Debye-Hückel theory.}\label{Sij_osmotic}
\end{figure}
According to the Kirkwood-Buff theory \cite{kirkwood1951statistical}, thermodynamic derivatives can be computed using radial distribution functions via the so-called Kirkwood-Buff integrals,
as $\lim_{R\to\infty} S_{ij}(R)$, with:
\begin{equation}
S_{ij}(R)=\delta_{ij}+\langle\rho_{salt}\rangle\int_{0}^R(g_{ij}(r)-1)4\pi r^2dr,
\end{equation}
where $g_{ij}(r)$ is a partial radial distribution function between ions of species $i$ and $j$. In the $N_{\text{water}}\mu_{\text{NaCl}}pT$ ensemble, $S_{ij}(R)$ gives the osmotic compressibility (Equation 1 in the main text). If $g_{ij}(r)\approx 1-\beta\frac{q_iq_j}{4\pi\epsilon_0\epsilon_s r}\exp(-\kappa_D r)$, following the Debye-Hückel approximation with the Debye screening parameter $\kappa_D$, then $S_{ij}(R)$ converges to 0.5 as $R\rightarrow\infty$. As already noted, the Debye-Hückel approximation for the compressibility reduces to that for the ideal gas for sufficiently dilute electrolytes.
Figure~\ref{Sij_osmotic} clearly shows that all three $S_{ij}(R)$ converge to the same value close to 0.5, satisfying the charge neutrality condition, as in Ref~\onlinecite{belloni2019non}.

%\end{suppinfo}
\newpage
\bibliography{manuscript}

%aipnum4-2.bst 2019-01-14 (MD) hand-edited version of apsrev4-1.bst
\begin{thebibliography}{57}%
\makeatletter
\providecommand \@ifxundefined [1]{%
 \@ifx{#1\undefined}
}%
\providecommand \@ifnum [1]{%
 \ifnum #1\expandafter \@firstoftwo
 \else \expandafter \@secondoftwo
 \fi
}%
\providecommand \@ifx [1]{%
 \ifx #1\expandafter \@firstoftwo
 \else \expandafter \@secondoftwo
 \fi
}%
\providecommand \natexlab [1]{#1}%
\providecommand \enquote  [1]{``#1''}%
\providecommand \bibnamefont  [1]{#1}%
\providecommand \bibfnamefont [1]{#1}%
\providecommand \citenamefont [1]{#1}%
\providecommand \href@noop [0]{\@secondoftwo}%
\providecommand \href [0]{\begingroup \@sanitize@url \@href}%
\providecommand \@href[1]{\@@startlink{#1}\@@href}%
\providecommand \@@href[1]{\endgroup#1\@@endlink}%
\providecommand \@sanitize@url [0]{\catcode `\\12\catcode `\$12\catcode
  `\&12\catcode `\#12\catcode `\^12\catcode `\_12\catcode `\%12\relax}%
\providecommand \@@startlink[1]{}%
\providecommand \@@endlink[0]{}%
\providecommand \url  [0]{\begingroup\@sanitize@url \@url }%
\providecommand \@url [1]{\endgroup\@href {#1}{\urlprefix }}%
\providecommand \urlprefix  [0]{URL }%
\providecommand \Eprint [0]{\href }%
\providecommand \doibase [0]{https://doi.org/}%
\providecommand \selectlanguage [0]{\@gobble}%
\providecommand \bibinfo  [0]{\@secondoftwo}%
\providecommand \bibfield  [0]{\@secondoftwo}%
\providecommand \translation [1]{[#1]}%
\providecommand \BibitemOpen [0]{}%
\providecommand \bibitemStop [0]{}%
\providecommand \bibitemNoStop [0]{.\EOS\space}%
\providecommand \EOS [0]{\spacefactor3000\relax}%
\providecommand \BibitemShut  [1]{\csname bibitem#1\endcsname}%
\let\auto@bib@innerbib\@empty
%</preamble>
\bibitem [{\citenamefont {Valeau}\ and\ \citenamefont
  {Cohen}(1980)}]{valeau1980primitive}%
  \BibitemOpen
  \bibfield  {author} {\bibinfo {author} {\bibfnamefont {J.}~\bibnamefont
  {Valeau}}\ and\ \bibinfo {author} {\bibfnamefont {L.}~\bibnamefont {Cohen}},\
  }\bibfield  {title} {\enquote {\bibinfo {title} {Primitive model electrolyte.
  i. grand canonical monte carlo computation},}\ }\href@noop {} {\bibfield
  {journal} {\bibinfo  {journal} {The Journal of Chemical Physics}\ }\textbf
  {\bibinfo {volume} {72}},\ \bibinfo {pages} {5935--5941} (\bibinfo {year}
  {1980})}\BibitemShut {NoStop}%
\bibitem [{\citenamefont {Belloni}(2019)}]{belloni2019non}%
  \BibitemOpen
  \bibfield  {author} {\bibinfo {author} {\bibfnamefont {L.}~\bibnamefont
  {Belloni}},\ }\bibfield  {title} {\enquote {\bibinfo {title} {Non-equilibrium
  hybrid insertion/extraction through the 4th dimension in grand-canonical
  simulation},}\ }\href@noop {} {\bibfield  {journal} {\bibinfo  {journal} {The
  Journal of Chemical Physics}\ }\textbf {\bibinfo {volume} {151}},\ \bibinfo
  {pages} {021101} (\bibinfo {year} {2019})}\BibitemShut {NoStop}%
\bibitem [{\citenamefont {Smit}(1995)}]{smit1995grand}%
  \BibitemOpen
  \bibfield  {author} {\bibinfo {author} {\bibfnamefont {B.}~\bibnamefont
  {Smit}},\ }\bibfield  {title} {\enquote {\bibinfo {title} {Grand canonical
  monte carlo simulations of chain molecules: adsorption isotherms of alkanes
  in zeolites},}\ }\href@noop {} {\bibfield  {journal} {\bibinfo  {journal}
  {Molecular Physics}\ }\textbf {\bibinfo {volume} {85}},\ \bibinfo {pages}
  {153--172} (\bibinfo {year} {1995})}\BibitemShut {NoStop}%
\bibitem [{\citenamefont {Stern}(2007)}]{stern2007molecular}%
  \BibitemOpen
  \bibfield  {author} {\bibinfo {author} {\bibfnamefont {H.~A.}\ \bibnamefont
  {Stern}},\ }\bibfield  {title} {\enquote {\bibinfo {title} {Molecular
  simulation with variable protonation states at constant p h},}\ }\href@noop
  {} {\bibfield  {journal} {\bibinfo  {journal} {The Journal of Chemical
  Physics}\ }\textbf {\bibinfo {volume} {126}},\ \bibinfo {pages} {04B627}
  (\bibinfo {year} {2007})}\BibitemShut {NoStop}%
\bibitem [{\citenamefont {Michael}\ \emph {et~al.}(2020)\citenamefont
  {Michael}, \citenamefont {Polydorides}, \citenamefont {Simonson},\ and\
  \citenamefont {Archontis}}]{michael2020hybrid}%
  \BibitemOpen
  \bibfield  {author} {\bibinfo {author} {\bibfnamefont {E.}~\bibnamefont
  {Michael}}, \bibinfo {author} {\bibfnamefont {S.}~\bibnamefont
  {Polydorides}}, \bibinfo {author} {\bibfnamefont {T.}~\bibnamefont
  {Simonson}},\ and\ \bibinfo {author} {\bibfnamefont {G.}~\bibnamefont
  {Archontis}},\ }\bibfield  {title} {\enquote {\bibinfo {title} {Hybrid mc/md
  for protein design},}\ }\href@noop {} {\bibfield  {journal} {\bibinfo
  {journal} {The Journal of Chemical Physics}\ }\textbf {\bibinfo {volume}
  {153}},\ \bibinfo {pages} {054113} (\bibinfo {year} {2020})}\BibitemShut
  {NoStop}%
\bibitem [{\citenamefont {Kirkwood}\ and\ \citenamefont
  {Buff}(1951)}]{kirkwood1951statistical}%
  \BibitemOpen
  \bibfield  {author} {\bibinfo {author} {\bibfnamefont {J.~G.}\ \bibnamefont
  {Kirkwood}}\ and\ \bibinfo {author} {\bibfnamefont {F.~P.}\ \bibnamefont
  {Buff}},\ }\bibfield  {title} {\enquote {\bibinfo {title} {The statistical
  mechanical theory of solutions. i},}\ }\href@noop {} {\bibfield  {journal}
  {\bibinfo  {journal} {The Journal of Chemical Physics}\ }\textbf {\bibinfo
  {volume} {19}},\ \bibinfo {pages} {774--777} (\bibinfo {year}
  {1951})}\BibitemShut {NoStop}%
\bibitem [{\citenamefont {Kusalik}\ and\ \citenamefont
  {Patey}(1987)}]{kusalik1987thermodynamic}%
  \BibitemOpen
  \bibfield  {author} {\bibinfo {author} {\bibfnamefont {P.~G.}\ \bibnamefont
  {Kusalik}}\ and\ \bibinfo {author} {\bibfnamefont {G.}~\bibnamefont
  {Patey}},\ }\bibfield  {title} {\enquote {\bibinfo {title} {The thermodynamic
  properties of electrolyte solutions: Some formal results},}\ }\href@noop {}
  {\bibfield  {journal} {\bibinfo  {journal} {The Journal of Chemical Physics}\
  }\textbf {\bibinfo {volume} {86}},\ \bibinfo {pages} {5110--5116} (\bibinfo
  {year} {1987})}\BibitemShut {NoStop}%
\bibitem [{\citenamefont {Cheng}(2022)}]{cheng2022computing}%
  \BibitemOpen
  \bibfield  {author} {\bibinfo {author} {\bibfnamefont {B.}~\bibnamefont
  {Cheng}},\ }\bibfield  {title} {\enquote {\bibinfo {title} {Computing
  chemical potentials of solutions from structure factors},}\ }\href@noop {}
  {\bibfield  {journal} {\bibinfo  {journal} {The Journal of Chemical Physics}\
  }\textbf {\bibinfo {volume} {157}},\ \bibinfo {pages} {121101--121106}
  (\bibinfo {year} {2022})}\BibitemShut {NoStop}%
\bibitem [{\citenamefont {Mezei}(1980)}]{mezei1980cavity}%
  \BibitemOpen
  \bibfield  {author} {\bibinfo {author} {\bibfnamefont {M.}~\bibnamefont
  {Mezei}},\ }\bibfield  {title} {\enquote {\bibinfo {title} {A cavity-biased
  (t, v, $\mu$) monte carlo method for the computer simulation of fluids},}\
  }\href@noop {} {\bibfield  {journal} {\bibinfo  {journal} {Molecular
  Physics}\ }\textbf {\bibinfo {volume} {40}},\ \bibinfo {pages} {901--906}
  (\bibinfo {year} {1980})}\BibitemShut {NoStop}%
\bibitem [{\citenamefont {Shelley}\ and\ \citenamefont
  {Patey}(1994)}]{shelley1994configuration}%
  \BibitemOpen
  \bibfield  {author} {\bibinfo {author} {\bibfnamefont {J.}~\bibnamefont
  {Shelley}}\ and\ \bibinfo {author} {\bibfnamefont {G.}~\bibnamefont
  {Patey}},\ }\bibfield  {title} {\enquote {\bibinfo {title} {A configuration
  bias monte carlo method for ionic solutions},}\ }\href@noop {} {\bibfield
  {journal} {\bibinfo  {journal} {The Journal of Chemical Physics}\ }\textbf
  {\bibinfo {volume} {100}},\ \bibinfo {pages} {8265--8270} (\bibinfo {year}
  {1994})}\BibitemShut {NoStop}%
\bibitem [{\citenamefont {Shelley}\ and\ \citenamefont
  {Patey}(1995)}]{shelley1995configuration}%
  \BibitemOpen
  \bibfield  {author} {\bibinfo {author} {\bibfnamefont {J.}~\bibnamefont
  {Shelley}}\ and\ \bibinfo {author} {\bibfnamefont {G.}~\bibnamefont
  {Patey}},\ }\bibfield  {title} {\enquote {\bibinfo {title} {A configuration
  bias monte carlo method for water},}\ }\href@noop {} {\bibfield  {journal}
  {\bibinfo  {journal} {The Journal of Chemical Physics}\ }\textbf {\bibinfo
  {volume} {102}},\ \bibinfo {pages} {7656--7663} (\bibinfo {year}
  {1995})}\BibitemShut {NoStop}%
\bibitem [{\citenamefont {Garberoglio}(2008)}]{garberoglio2008boltzmann}%
  \BibitemOpen
  \bibfield  {author} {\bibinfo {author} {\bibfnamefont {G.}~\bibnamefont
  {Garberoglio}},\ }\bibfield  {title} {\enquote {\bibinfo {title} {Boltzmann
  bias grand canonical monte carlo},}\ }\href@noop {} {\bibfield  {journal}
  {\bibinfo  {journal} {The Journal of Chemical Physics}\ }\textbf {\bibinfo
  {volume} {128}},\ \bibinfo {pages} {134109} (\bibinfo {year}
  {2008})}\BibitemShut {NoStop}%
\bibitem [{\citenamefont {Shi}\ and\ \citenamefont
  {Maginn}(2007)}]{shi2007continuous}%
  \BibitemOpen
  \bibfield  {author} {\bibinfo {author} {\bibfnamefont {W.}~\bibnamefont
  {Shi}}\ and\ \bibinfo {author} {\bibfnamefont {E.~J.}\ \bibnamefont
  {Maginn}},\ }\bibfield  {title} {\enquote {\bibinfo {title} {Continuous
  fractional component monte carlo: an adaptive biasing method for open system
  atomistic simulations},}\ }\href@noop {} {\bibfield  {journal} {\bibinfo
  {journal} {Journal of Chemical Theory and Computation}\ }\textbf {\bibinfo
  {volume} {3}},\ \bibinfo {pages} {1451--1463} (\bibinfo {year}
  {2007})}\BibitemShut {NoStop}%
\bibitem [{\citenamefont {Panagiotopoulos}(1989)}]{panagiotopoulos1989exact}%
  \BibitemOpen
  \bibfield  {author} {\bibinfo {author} {\bibfnamefont {A.}~\bibnamefont
  {Panagiotopoulos}},\ }\bibfield  {title} {\enquote {\bibinfo {title} {Exact
  calculations of fluid-phase equilibria by monte carlo simulation in a new
  statistical ensemble},}\ }\href@noop {} {\bibfield  {journal} {\bibinfo
  {journal} {International Journal of Thermophysics}\ }\textbf {\bibinfo
  {volume} {10}},\ \bibinfo {pages} {447--457} (\bibinfo {year}
  {1989})}\BibitemShut {NoStop}%
\bibitem [{\citenamefont {Soroush~Barhaghi}\ \emph {et~al.}(2018)\citenamefont
  {Soroush~Barhaghi}, \citenamefont {Torabi}, \citenamefont {Nejahi},
  \citenamefont {Schwiebert},\ and\ \citenamefont
  {Potoff}}]{soroush2018molecular}%
  \BibitemOpen
  \bibfield  {author} {\bibinfo {author} {\bibfnamefont {M.}~\bibnamefont
  {Soroush~Barhaghi}}, \bibinfo {author} {\bibfnamefont {K.}~\bibnamefont
  {Torabi}}, \bibinfo {author} {\bibfnamefont {Y.}~\bibnamefont {Nejahi}},
  \bibinfo {author} {\bibfnamefont {L.}~\bibnamefont {Schwiebert}},\ and\
  \bibinfo {author} {\bibfnamefont {J.~J.}\ \bibnamefont {Potoff}},\ }\bibfield
   {title} {\enquote {\bibinfo {title} {Molecular exchange monte carlo: A
  generalized method for identity exchanges in grand canonical monte carlo
  simulations},}\ }\href@noop {} {\bibfield  {journal} {\bibinfo  {journal}
  {The Journal of Chemical Physics}\ }\textbf {\bibinfo {volume} {149}},\
  \bibinfo {pages} {072318} (\bibinfo {year} {2018})}\BibitemShut {NoStop}%
\bibitem [{\citenamefont {Fathizadeh}\ and\ \citenamefont
  {Elber}(2018)}]{fathizadeh2018mixed}%
  \BibitemOpen
  \bibfield  {author} {\bibinfo {author} {\bibfnamefont {A.}~\bibnamefont
  {Fathizadeh}}\ and\ \bibinfo {author} {\bibfnamefont {R.}~\bibnamefont
  {Elber}},\ }\bibfield  {title} {\enquote {\bibinfo {title} {A mixed
  alchemical and equilibrium dynamics to simulate heterogeneous dense fluids:
  Illustrations for lennard-jones mixtures and phospholipid membranes},}\
  }\href@noop {} {\bibfield  {journal} {\bibinfo  {journal} {The Journal of
  Chemical Physics}\ }\textbf {\bibinfo {volume} {149}},\ \bibinfo {pages}
  {072325} (\bibinfo {year} {2018})}\BibitemShut {NoStop}%
\bibitem [{\citenamefont {Duane}\ \emph {et~al.}(1987)\citenamefont {Duane},
  \citenamefont {Kennedy}, \citenamefont {Pendleton},\ and\ \citenamefont
  {Roweth}}]{duane1987hybrid}%
  \BibitemOpen
  \bibfield  {author} {\bibinfo {author} {\bibfnamefont {S.}~\bibnamefont
  {Duane}}, \bibinfo {author} {\bibfnamefont {A.~D.}\ \bibnamefont {Kennedy}},
  \bibinfo {author} {\bibfnamefont {B.~J.}\ \bibnamefont {Pendleton}},\ and\
  \bibinfo {author} {\bibfnamefont {D.}~\bibnamefont {Roweth}},\ }\bibfield
  {title} {\enquote {\bibinfo {title} {Hybrid monte carlo},}\ }\href@noop {}
  {\bibfield  {journal} {\bibinfo  {journal} {Physics Letters B}\ }\textbf
  {\bibinfo {volume} {195}},\ \bibinfo {pages} {216--222} (\bibinfo {year}
  {1987})}\BibitemShut {NoStop}%
\bibitem [{\citenamefont {Mehlig}, \citenamefont {Heermann},\ and\
  \citenamefont {Forrest}(1992)}]{mehlig1992hybrid}%
  \BibitemOpen
  \bibfield  {author} {\bibinfo {author} {\bibfnamefont {B.}~\bibnamefont
  {Mehlig}}, \bibinfo {author} {\bibfnamefont {D.}~\bibnamefont {Heermann}},\
  and\ \bibinfo {author} {\bibfnamefont {B.}~\bibnamefont {Forrest}},\
  }\bibfield  {title} {\enquote {\bibinfo {title} {Hybrid monte carlo method
  for condensed-matter systems},}\ }\href@noop {} {\bibfield  {journal}
  {\bibinfo  {journal} {Physical Review B}\ }\textbf {\bibinfo {volume} {45}},\
  \bibinfo {pages} {679} (\bibinfo {year} {1992})}\BibitemShut {NoStop}%
\bibitem [{\citenamefont {Boinepalli}\ and\ \citenamefont
  {Attard}(2003)}]{boinepalli2003grand}%
  \BibitemOpen
  \bibfield  {author} {\bibinfo {author} {\bibfnamefont {S.}~\bibnamefont
  {Boinepalli}}\ and\ \bibinfo {author} {\bibfnamefont {P.}~\bibnamefont
  {Attard}},\ }\bibfield  {title} {\enquote {\bibinfo {title} {Grand canonical
  molecular dynamics},}\ }\href@noop {} {\bibfield  {journal} {\bibinfo
  {journal} {The Journal of Chemical Physics}\ }\textbf {\bibinfo {volume}
  {119}},\ \bibinfo {pages} {12769--12775} (\bibinfo {year}
  {2003})}\BibitemShut {NoStop}%
\bibitem [{\citenamefont {Chen}\ and\ \citenamefont
  {Roux}(2014)}]{chen2014efficient}%
  \BibitemOpen
  \bibfield  {author} {\bibinfo {author} {\bibfnamefont {Y.}~\bibnamefont
  {Chen}}\ and\ \bibinfo {author} {\bibfnamefont {B.}~\bibnamefont {Roux}},\
  }\bibfield  {title} {\enquote {\bibinfo {title} {Efficient hybrid
  non-equilibrium molecular dynamics-monte carlo simulations with symmetric
  momentum reversal},}\ }\href@noop {} {\bibfield  {journal} {\bibinfo
  {journal} {The Journal of Chemical Physics}\ }\textbf {\bibinfo {volume}
  {141}},\ \bibinfo {pages} {09B612\_1} (\bibinfo {year} {2014})}\BibitemShut
  {NoStop}%
\bibitem [{\citenamefont {Radak}\ \emph {et~al.}(2017)\citenamefont {Radak},
  \citenamefont {Chipot}, \citenamefont {Suh}, \citenamefont {Jo},
  \citenamefont {Jiang}, \citenamefont {Phillips}, \citenamefont {Schulten},\
  and\ \citenamefont {Roux}}]{radak2017constant}%
  \BibitemOpen
  \bibfield  {author} {\bibinfo {author} {\bibfnamefont {B.~K.}\ \bibnamefont
  {Radak}}, \bibinfo {author} {\bibfnamefont {C.}~\bibnamefont {Chipot}},
  \bibinfo {author} {\bibfnamefont {D.}~\bibnamefont {Suh}}, \bibinfo {author}
  {\bibfnamefont {S.}~\bibnamefont {Jo}}, \bibinfo {author} {\bibfnamefont
  {W.}~\bibnamefont {Jiang}}, \bibinfo {author} {\bibfnamefont {J.~C.}\
  \bibnamefont {Phillips}}, \bibinfo {author} {\bibfnamefont {K.}~\bibnamefont
  {Schulten}},\ and\ \bibinfo {author} {\bibfnamefont {B.}~\bibnamefont
  {Roux}},\ }\bibfield  {title} {\enquote {\bibinfo {title} {Constant-ph
  molecular dynamics simulations for large biomolecular systems},}\ }\href@noop
  {} {\bibfield  {journal} {\bibinfo  {journal} {Journal of Chemical Theory and
  Computation}\ }\textbf {\bibinfo {volume} {13}},\ \bibinfo {pages}
  {5933--5944} (\bibinfo {year} {2017})}\BibitemShut {NoStop}%
\bibitem [{\citenamefont {Ross}\ \emph {et~al.}(2018)\citenamefont {Ross},
  \citenamefont {Rustenburg}, \citenamefont {Grinaway}, \citenamefont {Fass},\
  and\ \citenamefont {Chodera}}]{ross2018biomolecular}%
  \BibitemOpen
  \bibfield  {author} {\bibinfo {author} {\bibfnamefont {G.~A.}\ \bibnamefont
  {Ross}}, \bibinfo {author} {\bibfnamefont {A.~S.}\ \bibnamefont
  {Rustenburg}}, \bibinfo {author} {\bibfnamefont {P.~B.}\ \bibnamefont
  {Grinaway}}, \bibinfo {author} {\bibfnamefont {J.}~\bibnamefont {Fass}},\
  and\ \bibinfo {author} {\bibfnamefont {J.~D.}\ \bibnamefont {Chodera}},\
  }\bibfield  {title} {\enquote {\bibinfo {title} {Biomolecular simulations
  under realistic macroscopic salt conditions},}\ }\href@noop {} {\bibfield
  {journal} {\bibinfo  {journal} {The Journal of Physical Chemistry B}\
  }\textbf {\bibinfo {volume} {122}},\ \bibinfo {pages} {5466--5486} (\bibinfo
  {year} {2018})}\BibitemShut {NoStop}%
\bibitem [{\citenamefont {Prokhorenko}\ \emph {et~al.}(2018)\citenamefont
  {Prokhorenko}, \citenamefont {Kalke}, \citenamefont {Nahas},\ and\
  \citenamefont {Bellaiche}}]{prokhorenko2018large}%
  \BibitemOpen
  \bibfield  {author} {\bibinfo {author} {\bibfnamefont {S.}~\bibnamefont
  {Prokhorenko}}, \bibinfo {author} {\bibfnamefont {K.}~\bibnamefont {Kalke}},
  \bibinfo {author} {\bibfnamefont {Y.}~\bibnamefont {Nahas}},\ and\ \bibinfo
  {author} {\bibfnamefont {L.}~\bibnamefont {Bellaiche}},\ }\bibfield  {title}
  {\enquote {\bibinfo {title} {Large scale hybrid monte carlo simulations for
  structure and property prediction},}\ }\href@noop {} {\bibfield  {journal}
  {\bibinfo  {journal} {npj Computational Materials}\ }\textbf {\bibinfo
  {volume} {4}},\ \bibinfo {pages} {1--7} (\bibinfo {year} {2018})}\BibitemShut
  {NoStop}%
\bibitem [{\citenamefont {Nilmeier}\ \emph {et~al.}(2011)\citenamefont
  {Nilmeier}, \citenamefont {Crooks}, \citenamefont {Minh},\ and\ \citenamefont
  {Chodera}}]{nilmeier2011nonequilibrium}%
  \BibitemOpen
  \bibfield  {author} {\bibinfo {author} {\bibfnamefont {J.~P.}\ \bibnamefont
  {Nilmeier}}, \bibinfo {author} {\bibfnamefont {G.~E.}\ \bibnamefont
  {Crooks}}, \bibinfo {author} {\bibfnamefont {D.~D.}\ \bibnamefont {Minh}},\
  and\ \bibinfo {author} {\bibfnamefont {J.~D.}\ \bibnamefont {Chodera}},\
  }\bibfield  {title} {\enquote {\bibinfo {title} {Nonequilibrium candidate
  monte carlo is an efficient tool for equilibrium simulation},}\ }\href@noop
  {} {\bibfield  {journal} {\bibinfo  {journal} {Proceedings of the National
  Academy of Sciences}\ }\textbf {\bibinfo {volume} {108}},\ \bibinfo {pages}
  {E1009--E1018} (\bibinfo {year} {2011})}\BibitemShut {NoStop}%
\bibitem [{\citenamefont {Schneck}\ and\ \citenamefont
  {Netz}(2011)}]{schneck2011simple}%
  \BibitemOpen
  \bibfield  {author} {\bibinfo {author} {\bibfnamefont {E.}~\bibnamefont
  {Schneck}}\ and\ \bibinfo {author} {\bibfnamefont {R.~R.}\ \bibnamefont
  {Netz}},\ }\bibfield  {title} {\enquote {\bibinfo {title} {From simple
  surface models to lipid membranes: Universal aspects of the hydration
  interaction from solvent-explicit simulations},}\ }\href@noop {} {\bibfield
  {journal} {\bibinfo  {journal} {Current opinion in colloid \& interface
  science}\ }\textbf {\bibinfo {volume} {16}},\ \bibinfo {pages} {607--611}
  (\bibinfo {year} {2011})}\BibitemShut {NoStop}%
\bibitem [{\citenamefont {Schneck}, \citenamefont {Sedlmeier},\ and\
  \citenamefont {Netz}(2012)}]{schneck2012hydration}%
  \BibitemOpen
  \bibfield  {author} {\bibinfo {author} {\bibfnamefont {E.}~\bibnamefont
  {Schneck}}, \bibinfo {author} {\bibfnamefont {F.}~\bibnamefont {Sedlmeier}},\
  and\ \bibinfo {author} {\bibfnamefont {R.~R.}\ \bibnamefont {Netz}},\
  }\bibfield  {title} {\enquote {\bibinfo {title} {Hydration repulsion between
  biomembranes results from an interplay of dehydration and depolarization},}\
  }\href@noop {} {\bibfield  {journal} {\bibinfo  {journal} {Proceedings of the
  National Academy of Sciences}\ }\textbf {\bibinfo {volume} {109}},\ \bibinfo
  {pages} {14405--14409} (\bibinfo {year} {2012})}\BibitemShut {NoStop}%
\bibitem [{\citenamefont {Schlaich}, \citenamefont {Knapp},\ and\ \citenamefont
  {Netz}(2016)}]{schlaich_water_2016}%
  \BibitemOpen
  \bibfield  {author} {\bibinfo {author} {\bibfnamefont {A.}~\bibnamefont
  {Schlaich}}, \bibinfo {author} {\bibfnamefont {E.~W.}\ \bibnamefont
  {Knapp}},\ and\ \bibinfo {author} {\bibfnamefont {R.~R.}\ \bibnamefont
  {Netz}},\ }\bibfield  {title} {\enquote {\bibinfo {title} {Water {Dielectric}
  {Effects} in {Planar} {Confinement}},}\ }\href
  {https://doi.org/10.1103/PhysRevLett.117.048001} {\bibfield  {journal}
  {\bibinfo  {journal} {Physical Review Letters}\ }\textbf {\bibinfo {volume}
  {117}},\ \bibinfo {pages} {048001} (\bibinfo {year} {2016})}\BibitemShut
  {NoStop}%
\bibitem [{\citenamefont {Izarra}\ \emph {et~al.}(2023)\citenamefont {Izarra},
  \citenamefont {Coudert}, \citenamefont {Fuchs},\ and\ \citenamefont
  {Boutin}}]{izarra2023alchemical}%
  \BibitemOpen
  \bibfield  {author} {\bibinfo {author} {\bibfnamefont {A.~d.}\ \bibnamefont
  {Izarra}}, \bibinfo {author} {\bibfnamefont {F.-X.}\ \bibnamefont {Coudert}},
  \bibinfo {author} {\bibfnamefont {A.~H.}\ \bibnamefont {Fuchs}},\ and\
  \bibinfo {author} {\bibfnamefont {A.}~\bibnamefont {Boutin}},\ }\bibfield
  {title} {\enquote {\bibinfo {title} {Alchemical osmostat for monte carlo
  simulation: Sampling aqueous electrolyte solution in open systems},}\
  }\href@noop {} {\bibfield  {journal} {\bibinfo  {journal} {The Journal of
  Physical Chemistry B}\ }\textbf {\bibinfo {volume} {127}},\ \bibinfo {pages}
  {766--776} (\bibinfo {year} {2023})}\BibitemShut {NoStop}%
\bibitem [{\citenamefont {Kurut}, \citenamefont {Fonseca},\ and\ \citenamefont
  {Boomsma}(2017)}]{kurut2017driving}%
  \BibitemOpen
  \bibfield  {author} {\bibinfo {author} {\bibfnamefont {A.}~\bibnamefont
  {Kurut}}, \bibinfo {author} {\bibfnamefont {R.}~\bibnamefont {Fonseca}},\
  and\ \bibinfo {author} {\bibfnamefont {W.}~\bibnamefont {Boomsma}},\
  }\bibfield  {title} {\enquote {\bibinfo {title} {Driving structural
  transitions in molecular simulations using the nonequilibrium candidate monte
  carlo},}\ }\href@noop {} {\bibfield  {journal} {\bibinfo  {journal} {The
  Journal of Physical Chemistry B}\ }\textbf {\bibinfo {volume} {122}},\
  \bibinfo {pages} {1195--1204} (\bibinfo {year} {2017})}\BibitemShut {NoStop}%
\bibitem [{\citenamefont {Suh}\ \emph {et~al.}(2018)\citenamefont {Suh},
  \citenamefont {Radak}, \citenamefont {Chipot},\ and\ \citenamefont
  {Roux}}]{suh2018enhanced}%
  \BibitemOpen
  \bibfield  {author} {\bibinfo {author} {\bibfnamefont {D.}~\bibnamefont
  {Suh}}, \bibinfo {author} {\bibfnamefont {B.~K.}\ \bibnamefont {Radak}},
  \bibinfo {author} {\bibfnamefont {C.}~\bibnamefont {Chipot}},\ and\ \bibinfo
  {author} {\bibfnamefont {B.}~\bibnamefont {Roux}},\ }\bibfield  {title}
  {\enquote {\bibinfo {title} {Enhanced configurational sampling with hybrid
  non-equilibrium molecular dynamics--monte carlo propagator},}\ }\href@noop {}
  {\bibfield  {journal} {\bibinfo  {journal} {The Journal of Chemical Physics}\
  }\textbf {\bibinfo {volume} {148}},\ \bibinfo {pages} {014101} (\bibinfo
  {year} {2018})}\BibitemShut {NoStop}%
\bibitem [{\citenamefont {Sasmal}\ \emph {et~al.}(2020)\citenamefont {Sasmal},
  \citenamefont {Gill}, \citenamefont {Lim},\ and\ \citenamefont
  {Mobley}}]{sasmal2020sampling}%
  \BibitemOpen
  \bibfield  {author} {\bibinfo {author} {\bibfnamefont {S.}~\bibnamefont
  {Sasmal}}, \bibinfo {author} {\bibfnamefont {S.~C.}\ \bibnamefont {Gill}},
  \bibinfo {author} {\bibfnamefont {N.~M.}\ \bibnamefont {Lim}},\ and\ \bibinfo
  {author} {\bibfnamefont {D.~L.}\ \bibnamefont {Mobley}},\ }\bibfield  {title}
  {\enquote {\bibinfo {title} {Sampling conformational changes of bound ligands
  using nonequilibrium candidate monte carlo and molecular dynamics},}\
  }\href@noop {} {\bibfield  {journal} {\bibinfo  {journal} {Journal of
  Chemical Theory and Computation}\ }\textbf {\bibinfo {volume} {16}},\
  \bibinfo {pages} {1854--1865} (\bibinfo {year} {2020})}\BibitemShut {NoStop}%
\bibitem [{\citenamefont {Voter}(1997)}]{voter1997hyperdynamics}%
  \BibitemOpen
  \bibfield  {author} {\bibinfo {author} {\bibfnamefont {A.~F.}\ \bibnamefont
  {Voter}},\ }\bibfield  {title} {\enquote {\bibinfo {title} {Hyperdynamics:
  Accelerated molecular dynamics of infrequent events},}\ }\href@noop {}
  {\bibfield  {journal} {\bibinfo  {journal} {Physical Review Letters}\
  }\textbf {\bibinfo {volume} {78}},\ \bibinfo {pages} {3908} (\bibinfo {year}
  {1997})}\BibitemShut {NoStop}%
\bibitem [{\citenamefont {Wang}, \citenamefont {Friesner},\ and\ \citenamefont
  {Berne}(2011)}]{wang2011replica}%
  \BibitemOpen
  \bibfield  {author} {\bibinfo {author} {\bibfnamefont {L.}~\bibnamefont
  {Wang}}, \bibinfo {author} {\bibfnamefont {R.~A.}\ \bibnamefont {Friesner}},\
  and\ \bibinfo {author} {\bibfnamefont {B.}~\bibnamefont {Berne}},\ }\bibfield
   {title} {\enquote {\bibinfo {title} {Replica exchange with solute scaling: a
  more efficient version of replica exchange with solute tempering (rest2)},}\
  }\href@noop {} {\bibfield  {journal} {\bibinfo  {journal} {The Journal of
  Physical Chemistry B}\ }\textbf {\bibinfo {volume} {115}},\ \bibinfo {pages}
  {9431--9438} (\bibinfo {year} {2011})}\BibitemShut {NoStop}%
\bibitem [{\citenamefont {Plimpton}(1995)}]{plimpton1995fast}%
  \BibitemOpen
  \bibfield  {author} {\bibinfo {author} {\bibfnamefont {S.}~\bibnamefont
  {Plimpton}},\ }\bibfield  {title} {\enquote {\bibinfo {title} {Fast parallel
  algorithms for short-range molecular dynamics},}\ }\href@noop {} {\bibfield
  {journal} {\bibinfo  {journal} {Journal of Computational Physics}\ }\textbf
  {\bibinfo {volume} {117}},\ \bibinfo {pages} {1--19} (\bibinfo {year}
  {1995})}\BibitemShut {NoStop}%
\bibitem [{\citenamefont {Guo}, \citenamefont {Haji-Akbari},\ and\
  \citenamefont {Palmer}(2018)}]{guo2018hybrid}%
  \BibitemOpen
  \bibfield  {author} {\bibinfo {author} {\bibfnamefont {J.}~\bibnamefont
  {Guo}}, \bibinfo {author} {\bibfnamefont {A.}~\bibnamefont {Haji-Akbari}},\
  and\ \bibinfo {author} {\bibfnamefont {J.~C.}\ \bibnamefont {Palmer}},\
  }\bibfield  {title} {\enquote {\bibinfo {title} {Hybrid monte carlo with
  lammps},}\ }\href@noop {} {\bibfield  {journal} {\bibinfo  {journal} {Journal
  of Theoretical and Computational Chemistry}\ }\textbf {\bibinfo {volume}
  {17}},\ \bibinfo {pages} {1840002} (\bibinfo {year} {2018})}\BibitemShut
  {NoStop}%
\bibitem [{\citenamefont {Palmer}\ \emph {et~al.}(2018)\citenamefont {Palmer},
  \citenamefont {Haji-Akbari}, \citenamefont {Singh}, \citenamefont {Martelli},
  \citenamefont {Car}, \citenamefont {Panagiotopoulos},\ and\ \citenamefont
  {Debenedetti}}]{palmer2018comment}%
  \BibitemOpen
  \bibfield  {author} {\bibinfo {author} {\bibfnamefont {J.~C.}\ \bibnamefont
  {Palmer}}, \bibinfo {author} {\bibfnamefont {A.}~\bibnamefont {Haji-Akbari}},
  \bibinfo {author} {\bibfnamefont {R.~S.}\ \bibnamefont {Singh}}, \bibinfo
  {author} {\bibfnamefont {F.}~\bibnamefont {Martelli}}, \bibinfo {author}
  {\bibfnamefont {R.}~\bibnamefont {Car}}, \bibinfo {author} {\bibfnamefont
  {A.~Z.}\ \bibnamefont {Panagiotopoulos}},\ and\ \bibinfo {author}
  {\bibfnamefont {P.~G.}\ \bibnamefont {Debenedetti}},\ }\bibfield  {title}
  {\enquote {\bibinfo {title} {Comment on “the putative liquid-liquid
  transition is a liquid-solid transition in atomistic models of water”},}\
  }\href@noop {} {\bibfield  {journal} {\bibinfo  {journal} {The Journal of
  Chemical Physics}\ }\textbf {\bibinfo {volume} {148}},\ \bibinfo {pages}
  {137101} (\bibinfo {year} {2018})}\BibitemShut {NoStop}%
\bibitem [{\citenamefont {Miller~III}\ \emph {et~al.}(2002)\citenamefont
  {Miller~III}, \citenamefont {Eleftheriou}, \citenamefont {Pattnaik},
  \citenamefont {Ndirango}, \citenamefont {Newns},\ and\ \citenamefont
  {Martyna}}]{miller2002symplectic}%
  \BibitemOpen
  \bibfield  {author} {\bibinfo {author} {\bibfnamefont {T.~F.}\ \bibnamefont
  {Miller~III}}, \bibinfo {author} {\bibfnamefont {M.}~\bibnamefont
  {Eleftheriou}}, \bibinfo {author} {\bibfnamefont {P.}~\bibnamefont
  {Pattnaik}}, \bibinfo {author} {\bibfnamefont {A.}~\bibnamefont {Ndirango}},
  \bibinfo {author} {\bibfnamefont {D.}~\bibnamefont {Newns}},\ and\ \bibinfo
  {author} {\bibfnamefont {G.}~\bibnamefont {Martyna}},\ }\bibfield  {title}
  {\enquote {\bibinfo {title} {Symplectic quaternion scheme for biophysical
  molecular dynamics},}\ }\href@noop {} {\bibfield  {journal} {\bibinfo
  {journal} {The Journal of Chemical Physics}\ }\textbf {\bibinfo {volume}
  {116}},\ \bibinfo {pages} {8649--8659} (\bibinfo {year} {2002})}\BibitemShut
  {NoStop}%
\bibitem [{\citenamefont {Frenkel}\ and\ \citenamefont
  {Smit}(2001)}]{frenkel2001understanding}%
  \BibitemOpen
  \bibfield  {author} {\bibinfo {author} {\bibfnamefont {D.}~\bibnamefont
  {Frenkel}}\ and\ \bibinfo {author} {\bibfnamefont {B.}~\bibnamefont {Smit}},\
  }\href@noop {} {\emph {\bibinfo {title} {Understanding molecular simulation:
  from algorithms to applications}}},\ Vol.~\bibinfo {volume} {1}\ (\bibinfo
  {publisher} {Elsevier},\ \bibinfo {year} {2001})\BibitemShut {NoStop}%
\bibitem [{\citenamefont {Inagaki}\ and\ \citenamefont
  {Saito}(2022)}]{inagaki2022hybrid}%
  \BibitemOpen
  \bibfield  {author} {\bibinfo {author} {\bibfnamefont {T.}~\bibnamefont
  {Inagaki}}\ and\ \bibinfo {author} {\bibfnamefont {S.}~\bibnamefont
  {Saito}},\ }\bibfield  {title} {\enquote {\bibinfo {title} {Hybrid monte
  carlo method with potential scaling for sampling from the canonical
  multimodal distribution and imitating the relaxation process},}\ }\href@noop
  {} {\bibfield  {journal} {\bibinfo  {journal} {The Journal of Chemical
  Physics}\ }\textbf {\bibinfo {volume} {156}},\ \bibinfo {pages} {104111}
  (\bibinfo {year} {2022})}\BibitemShut {NoStop}%
\bibitem [{\citenamefont {Stillinger~Jr}\ and\ \citenamefont
  {Lovett}(1968{\natexlab{a}})}]{stillinger1968ion}%
  \BibitemOpen
  \bibfield  {author} {\bibinfo {author} {\bibfnamefont {F.~H.}\ \bibnamefont
  {Stillinger~Jr}}\ and\ \bibinfo {author} {\bibfnamefont {R.}~\bibnamefont
  {Lovett}},\ }\bibfield  {title} {\enquote {\bibinfo {title} {Ion-pair theory
  of concentrated electrolytes. i. basic concepts},}\ }\href@noop {} {\bibfield
   {journal} {\bibinfo  {journal} {The Journal of Chemical Physics}\ }\textbf
  {\bibinfo {volume} {48}},\ \bibinfo {pages} {3858--3868} (\bibinfo {year}
  {1968}{\natexlab{a}})}\BibitemShut {NoStop}%
\bibitem [{\citenamefont {Stillinger~Jr}\ and\ \citenamefont
  {Lovett}(1968{\natexlab{b}})}]{stillinger1968general}%
  \BibitemOpen
  \bibfield  {author} {\bibinfo {author} {\bibfnamefont {F.~H.}\ \bibnamefont
  {Stillinger~Jr}}\ and\ \bibinfo {author} {\bibfnamefont {R.}~\bibnamefont
  {Lovett}},\ }\bibfield  {title} {\enquote {\bibinfo {title} {General
  restriction on the distribution of ions in electrolytes},}\ }\href@noop {}
  {\bibfield  {journal} {\bibinfo  {journal} {The Journal of Chemical Physics}\
  }\textbf {\bibinfo {volume} {49}},\ \bibinfo {pages} {1991--1994} (\bibinfo
  {year} {1968}{\natexlab{b}})}\BibitemShut {NoStop}%
\bibitem [{\citenamefont {Rosenbluth}\ and\ \citenamefont
  {Rosenbluth}(1955)}]{rosenbluth1955monte}%
  \BibitemOpen
  \bibfield  {author} {\bibinfo {author} {\bibfnamefont {M.~N.}\ \bibnamefont
  {Rosenbluth}}\ and\ \bibinfo {author} {\bibfnamefont {A.~W.}\ \bibnamefont
  {Rosenbluth}},\ }\bibfield  {title} {\enquote {\bibinfo {title} {Monte carlo
  calculation of the average extension of molecular chains},}\ }\href@noop {}
  {\bibfield  {journal} {\bibinfo  {journal} {The Journal of Chemical Physics}\
  }\textbf {\bibinfo {volume} {23}},\ \bibinfo {pages} {356--359} (\bibinfo
  {year} {1955})}\BibitemShut {NoStop}%
\bibitem [{\citenamefont {Joly}\ \emph {et~al.}(2006)\citenamefont {Joly},
  \citenamefont {Ybert}, \citenamefont {Trizac},\ and\ \citenamefont
  {Bocquet}}]{joly2006liquid}%
  \BibitemOpen
  \bibfield  {author} {\bibinfo {author} {\bibfnamefont {L.}~\bibnamefont
  {Joly}}, \bibinfo {author} {\bibfnamefont {C.}~\bibnamefont {Ybert}},
  \bibinfo {author} {\bibfnamefont {E.}~\bibnamefont {Trizac}},\ and\ \bibinfo
  {author} {\bibfnamefont {L.}~\bibnamefont {Bocquet}},\ }\bibfield  {title}
  {\enquote {\bibinfo {title} {Liquid friction on charged surfaces: From
  hydrodynamic slippage to electrokinetics},}\ }\href@noop {} {\bibfield
  {journal} {\bibinfo  {journal} {The Journal of Chemical Physics}\ }\textbf
  {\bibinfo {volume} {125}},\ \bibinfo {pages} {204716} (\bibinfo {year}
  {2006})}\BibitemShut {NoStop}%
\bibitem [{\citenamefont {Scalfi}, \citenamefont {Coasne},\ and\ \citenamefont
  {Rotenberg}(2021)}]{scalfi2021gibbs}%
  \BibitemOpen
  \bibfield  {author} {\bibinfo {author} {\bibfnamefont {L.}~\bibnamefont
  {Scalfi}}, \bibinfo {author} {\bibfnamefont {B.}~\bibnamefont {Coasne}},\
  and\ \bibinfo {author} {\bibfnamefont {B.}~\bibnamefont {Rotenberg}},\
  }\bibfield  {title} {\enquote {\bibinfo {title} {On the gibbs--thomson
  equation for the crystallization of confined fluids},}\ }\href@noop {}
  {\bibfield  {journal} {\bibinfo  {journal} {The Journal of Chemical Physics}\
  }\textbf {\bibinfo {volume} {154}},\ \bibinfo {pages} {114711} (\bibinfo
  {year} {2021})}\BibitemShut {NoStop}%
\bibitem [{\citenamefont {Berendsen}, \citenamefont {Grigera},\ and\
  \citenamefont {Straatsma}(1987)}]{berendsen1987missing}%
  \BibitemOpen
  \bibfield  {author} {\bibinfo {author} {\bibfnamefont {H.}~\bibnamefont
  {Berendsen}}, \bibinfo {author} {\bibfnamefont {J.}~\bibnamefont {Grigera}},\
  and\ \bibinfo {author} {\bibfnamefont {T.}~\bibnamefont {Straatsma}},\
  }\bibfield  {title} {\enquote {\bibinfo {title} {The missing term in
  effective pair potentials},}\ }\href@noop {} {\bibfield  {journal} {\bibinfo
  {journal} {Journal of Physical Chemistry}\ }\textbf {\bibinfo {volume}
  {91}},\ \bibinfo {pages} {6269--6271} (\bibinfo {year} {1987})}\BibitemShut
  {NoStop}%
\bibitem [{\citenamefont {Mou{\v{c}}ka}, \citenamefont {Nezbeda},\ and\
  \citenamefont {Smith}(2013)}]{mouvcka2013molecular}%
  \BibitemOpen
  \bibfield  {author} {\bibinfo {author} {\bibfnamefont {F.}~\bibnamefont
  {Mou{\v{c}}ka}}, \bibinfo {author} {\bibfnamefont {I.}~\bibnamefont
  {Nezbeda}},\ and\ \bibinfo {author} {\bibfnamefont {W.~R.}\ \bibnamefont
  {Smith}},\ }\bibfield  {title} {\enquote {\bibinfo {title} {Molecular
  simulation of aqueous electrolytes: Water chemical potential results and
  gibbs-duhem equation consistency tests},}\ }\href@noop {} {\bibfield
  {journal} {\bibinfo  {journal} {The Journal of Chemical Physics}\ }\textbf
  {\bibinfo {volume} {139}},\ \bibinfo {pages} {124505} (\bibinfo {year}
  {2013})}\BibitemShut {NoStop}%
\bibitem [{\citenamefont {Mester}\ and\ \citenamefont
  {Panagiotopoulos}(2015)}]{mester2015mean}%
  \BibitemOpen
  \bibfield  {author} {\bibinfo {author} {\bibfnamefont {Z.}~\bibnamefont
  {Mester}}\ and\ \bibinfo {author} {\bibfnamefont {A.~Z.}\ \bibnamefont
  {Panagiotopoulos}},\ }\bibfield  {title} {\enquote {\bibinfo {title} {Mean
  ionic activity coefficients in aqueous nacl solutions from molecular dynamics
  simulations},}\ }\href@noop {} {\bibfield  {journal} {\bibinfo  {journal}
  {The Journal of Chemical Physics}\ }\textbf {\bibinfo {volume} {142}},\
  \bibinfo {pages} {044507} (\bibinfo {year} {2015})}\BibitemShut {NoStop}%
\bibitem [{\citenamefont {Andersen}(1983)}]{andersen1983rattle}%
  \BibitemOpen
  \bibfield  {author} {\bibinfo {author} {\bibfnamefont {H.~C.}\ \bibnamefont
  {Andersen}},\ }\bibfield  {title} {\enquote {\bibinfo {title} {Rattle: A
  “velocity” version of the shake algorithm for molecular dynamics
  calculations},}\ }\href@noop {} {\bibfield  {journal} {\bibinfo  {journal}
  {Journal of Computational Physics}\ }\textbf {\bibinfo {volume} {52}},\
  \bibinfo {pages} {24--34} (\bibinfo {year} {1983})}\BibitemShut {NoStop}%
\bibitem [{\citenamefont {Widom}(1963)}]{widom1963some}%
  \BibitemOpen
  \bibfield  {author} {\bibinfo {author} {\bibfnamefont {B.}~\bibnamefont
  {Widom}},\ }\bibfield  {title} {\enquote {\bibinfo {title} {Some topics in
  the theory of fluids},}\ }\href@noop {} {\bibfield  {journal} {\bibinfo
  {journal} {The Journal of Chemical Physics}\ }\textbf {\bibinfo {volume}
  {39}},\ \bibinfo {pages} {2808--2812} (\bibinfo {year} {1963})}\BibitemShut
  {NoStop}%
\bibitem [{\citenamefont {Crooks}(1999)}]{crooks1999entropy}%
  \BibitemOpen
  \bibfield  {author} {\bibinfo {author} {\bibfnamefont {G.~E.}\ \bibnamefont
  {Crooks}},\ }\bibfield  {title} {\enquote {\bibinfo {title} {Entropy
  production fluctuation theorem and the nonequilibrium work relation for free
  energy differences},}\ }\href@noop {} {\bibfield  {journal} {\bibinfo
  {journal} {Physical Review E}\ }\textbf {\bibinfo {volume} {60}},\ \bibinfo
  {pages} {2721} (\bibinfo {year} {1999})}\BibitemShut {NoStop}%
\bibitem [{\citenamefont {Belloni}(2018)}]{belloni2018finite}%
  \BibitemOpen
  \bibfield  {author} {\bibinfo {author} {\bibfnamefont {L.}~\bibnamefont
  {Belloni}},\ }\bibfield  {title} {\enquote {\bibinfo {title} {Finite-size
  corrections in numerical simulation of liquid water},}\ }\href@noop {}
  {\bibfield  {journal} {\bibinfo  {journal} {The Journal of Chemical Physics}\
  }\textbf {\bibinfo {volume} {149}} (\bibinfo {year} {2018})}\BibitemShut
  {NoStop}%
\bibitem [{che()}]{chempotconv}%
  \BibitemOpen
  \href@noop {} {\enquote {\bibinfo {title} {The difference in definition of
  ideal part of chemical potential in this work and
  ref.~\onlinecite{mester2015mean} requires a conversion of chemical potential.
  firstly, we set the thermal de broglie length unity as described below
  eq.~\ref{metro_del} for chemical potentials of both water and salt. secondly,
  the chemical potential of water in this work does not include
  $\mu^0_{\text{h$_2$o}}$ the standard state free energy of formation of water
  in the gas phase (see eq. 30 in ref.~\onlinecite{mester2015mean}). likewise,
  the chemical potential of salt in this work does not include
  $\mu^0_{\text{na$^+$}}$ and $\mu^0_{\text{cl$^-$}}$ (see eq. 17 in
  ref.~\onlinecite{mester2015mean}). one can obtain the chemical potential as
  in this work using the values of the excess chemical potential in si of
  ref.~\onlinecite{mester2015mean} and our definition of ideal gas contribution
  to the chemical potential.}}\ }\BibitemShut {NoStop}%
\bibitem [{\citenamefont {McQuarrie}(2000)}]{mcquarrie2000statistical}%
  \BibitemOpen
  \bibfield  {author} {\bibinfo {author} {\bibfnamefont {D.~A.}\ \bibnamefont
  {McQuarrie}},\ }\href@noop {} {\emph {\bibinfo {title} {Statistical
  mechanics}}}\ (\bibinfo  {publisher} {Sterling Publishing Company},\ \bibinfo
  {year} {2000})\BibitemShut {NoStop}%
\bibitem [{\citenamefont {Dellago}\ and\ \citenamefont
  {Hummer}(2013)}]{dellago2013computing}%
  \BibitemOpen
  \bibfield  {author} {\bibinfo {author} {\bibfnamefont {C.}~\bibnamefont
  {Dellago}}\ and\ \bibinfo {author} {\bibfnamefont {G.}~\bibnamefont
  {Hummer}},\ }\bibfield  {title} {\enquote {\bibinfo {title} {Computing
  equilibrium free energies using non-equilibrium molecular dynamics},}\
  }\href@noop {} {\bibfield  {journal} {\bibinfo  {journal} {Entropy}\ }\textbf
  {\bibinfo {volume} {16}},\ \bibinfo {pages} {41--61} (\bibinfo {year}
  {2013})}\BibitemShut {NoStop}%
\bibitem [{\citenamefont {Collin}\ \emph {et~al.}(2005)\citenamefont {Collin},
  \citenamefont {Ritort}, \citenamefont {Jarzynski}, \citenamefont {Smith},
  \citenamefont {Tinoco},\ and\ \citenamefont
  {Bustamante}}]{collin2005verification}%
  \BibitemOpen
  \bibfield  {author} {\bibinfo {author} {\bibfnamefont {D.}~\bibnamefont
  {Collin}}, \bibinfo {author} {\bibfnamefont {F.}~\bibnamefont {Ritort}},
  \bibinfo {author} {\bibfnamefont {C.}~\bibnamefont {Jarzynski}}, \bibinfo
  {author} {\bibfnamefont {S.~B.}\ \bibnamefont {Smith}}, \bibinfo {author}
  {\bibfnamefont {I.}~\bibnamefont {Tinoco}},\ and\ \bibinfo {author}
  {\bibfnamefont {C.}~\bibnamefont {Bustamante}},\ }\bibfield  {title}
  {\enquote {\bibinfo {title} {Verification of the crooks fluctuation theorem
  and recovery of rna folding free energies},}\ }\href@noop {} {\bibfield
  {journal} {\bibinfo  {journal} {Nature}\ }\textbf {\bibinfo {volume} {437}},\
  \bibinfo {pages} {231--234} (\bibinfo {year} {2005})}\BibitemShut {NoStop}%
\bibitem [{\citenamefont {Bennett}(1976)}]{bennett1976efficient}%
  \BibitemOpen
  \bibfield  {author} {\bibinfo {author} {\bibfnamefont {C.~H.}\ \bibnamefont
  {Bennett}},\ }\bibfield  {title} {\enquote {\bibinfo {title} {Efficient
  estimation of free energy differences from monte carlo data},}\ }\href@noop
  {} {\bibfield  {journal} {\bibinfo  {journal} {Journal of Computational
  Physics}\ }\textbf {\bibinfo {volume} {22}},\ \bibinfo {pages} {245--268}
  (\bibinfo {year} {1976})}\BibitemShut {NoStop}%
\bibitem [{\citenamefont {Shirts}\ \emph {et~al.}(2003)\citenamefont {Shirts},
  \citenamefont {Bair}, \citenamefont {Hooker},\ and\ \citenamefont
  {Pande}}]{shirts2003equilibrium}%
  \BibitemOpen
  \bibfield  {author} {\bibinfo {author} {\bibfnamefont {M.~R.}\ \bibnamefont
  {Shirts}}, \bibinfo {author} {\bibfnamefont {E.}~\bibnamefont {Bair}},
  \bibinfo {author} {\bibfnamefont {G.}~\bibnamefont {Hooker}},\ and\ \bibinfo
  {author} {\bibfnamefont {V.~S.}\ \bibnamefont {Pande}},\ }\bibfield  {title}
  {\enquote {\bibinfo {title} {Equilibrium free energies from nonequilibrium
  measurements using maximum-likelihood methods},}\ }\href@noop {} {\bibfield
  {journal} {\bibinfo  {journal} {Physical Review Letters}\ }\textbf {\bibinfo
  {volume} {91}},\ \bibinfo {pages} {140601} (\bibinfo {year}
  {2003})}\BibitemShut {NoStop}%
\end{thebibliography}%
\end{document}